\documentclass{article} 
\usepackage[final]{colm2026_conference}

\usepackage{microtype}
\usepackage{url}
\usepackage{booktabs}

\usepackage{booktabs,longtable,array,pifont,xcolor}

\renewcommand{\arraystretch}{1.08}
\usepackage{hyperref}
\usepackage{graphicx}
\usepackage{subcaption}
\usepackage{amsmath}
\usepackage{cleveref}
\usepackage{siunitx}
\usepackage{array,tabularx,booktabs}
\usepackage{multirow}
\usepackage{wrapfig}
\usepackage{caption}
\usepackage[figuresleft]{rotating}
\usepackage{float}

\usepackage[table]{xcolor}
\usepackage[most]{tcolorbox}
\tcbuselibrary{breakable,skins,listings}
\usepackage{placeins}
\usepackage{xspace}
\usepackage{fontawesome5}
\usepackage{enumitem}

\definecolor{takeawaybg}{HTML}{FFE0F9}
\definecolor{takeawayborder}{HTML}{290020}

\newtcolorbox{takeawaybox}{
    enhanced,
    colback=takeawaybg!45!white,
    colframe=takeawayborder,
    boxrule=1pt,
    arc=0.6mm, 
    left=2mm,
    right=2mm,
    top=1mm,
    bottom=1mm
}

\newtcolorbox{takeawaybox2}{
    enhanced,
    colback=takeawaybg!45!white,
    colframe=takeawayborder,
    boxrule=1pt,
    arc=0.6mm, 
    left=2mm,
    right=2mm,
    top=1mm,
    bottom=1mm,
    breakable
}
\newcolumntype{L}[1]{>{\raggedright\arraybackslash}p{#1}}
\newcolumntype{Y}{>{\raggedright\arraybackslash}X}

\newcommand{\huggingface}{
  \raisebox{-1.5pt}{
    \includegraphics[height=1.05em]{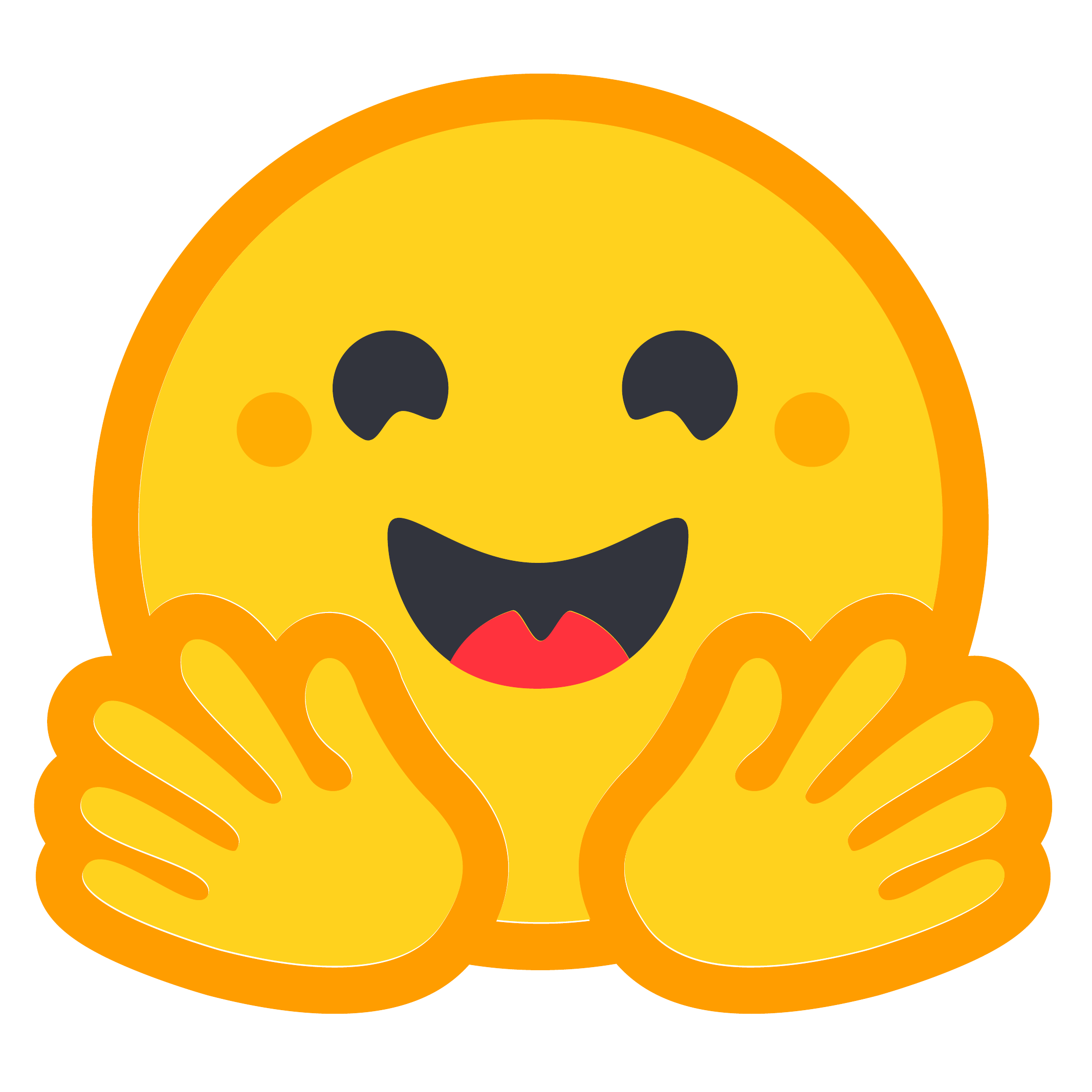}
  }\xspace
}

\newcommand{\github}{
  \raisebox{-1.5pt}{
    \includegraphics[height=1.05em]{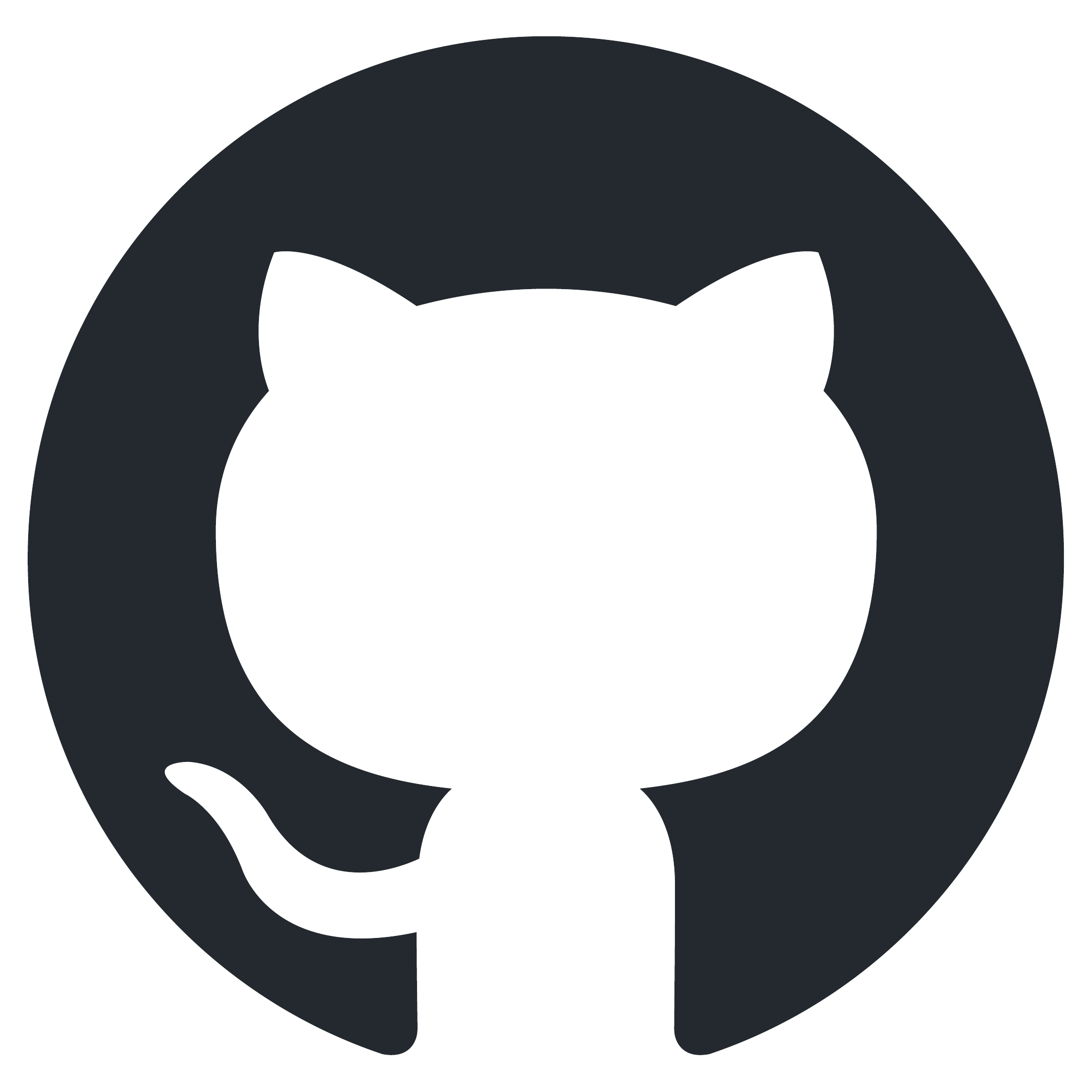}
  }\xspace
}

\usepackage{lineno}

\definecolor{darkblue}{rgb}{0, 0, 0.5}
\hypersetup{colorlinks=true, citecolor=darkblue, linkcolor=darkblue, urlcolor=darkblue}

\title{Process-Oriented Evaluation of AI-Assisted Scientific Writing}

\author{Patrick Queiroz Da Silva$^{\clubsuit}$ \quad Sanchaita Hazra$^{\heartsuit}$ \quad \textbf{Doeun Lee}$^{\clubsuit}$ \quad  \\ \textbf{Sachin Kumar}$^{\clubsuit}$ \quad \textbf{Bodhisattwa Prasad Majumder}$^{\diamondsuit}$ \\
$^\clubsuit$The Ohio State University, Columbus, OH \\
$^\heartsuit$University of Utah, Salt Lake City, UT \\
$^\diamondsuit$Allen Institute for AI, Seattle, WA \\\\
\href{https://github.com/skai-research/process-revision-sci-writing}
    {\github\ Code}, 
\href{https://huggingface.co/datasets/patrickqdasilva/process-revision-sci-write}
    {\huggingface\ Data} \\
Emails: \texttt{\small dasilva.30@osu.edu, sanchaita.hazra@utah.edu} 
} 

\begin{document}

\ifcolmsubmission
\linenumbers
\fi

\maketitle

\begin{abstract}
Bad writing hinders the publication of science. The role of artificial intelligence (AI) in generating and editing scientific texts remains unsettled. Abstracts serve as the critical gateway to scientific manuscripts, often shaping readers' interest. We inspect how individuals revise AI-generated abstracts compared to human-authored abstracts when incentivized to communicate scientific content. Using 869 keystroke-level edit logs with 240k total edits, we construct behavioral labels and measure linguistic properties of edit bursts to investigate the edit trajectories. AI abstracts exhibit higher sentence-level agency, whereas human-authored abstracts outperform in global coherence, even with edits. Experts engage in stigmatic behavior, switching their strategy from predominantly restructuring to substitution when AI source is disclosed. Language Models (LMs) improve edit outcomes through a mix of local and global features, but still actively struggle with global coherence. Both humans and LMs often target the weakest sections of abstracts, but fail to improve stronger areas. Our large-scale process-oriented evaluation highlights the perks and pitfalls of both human and LM editing processes as machine-generated texts emerge in scientific communication.
\end{abstract}

\section{Introduction}

Scientific texts are increasingly being shaped by Language Models (LMs). 
By helping reduce time and (often) effort required for drafting and revision, LMs are being used by researchers to  accelerate key writing tasks 
\citep{liang2025quantifying, kusumegi2025scientific}.
AI-assisted scientific writing is often perceived as more readable, which can boost writers' confidence and lead to fewer edits \citep{markowitz2024complexity, mcminn2025using, Hazra_2026}.
To boot, studies find that AI-generated scientific abstracts appear credible and, with minor editing, are often accepted by reviewers \citep{gao2023comparing, Holland2024, Hazra_2026}.  
At the same time, other work shows AI-generated scientific texts to lack in detail and insight \citep{tang2023evaluating, Hadan2024_dg, Peters2025_hp}.

These diverging observations point to a key missing piece: how authors revise AI-generated scientific text. 
Revision and refinement lie at the heart of the writing process, offering a rich window into how authors strategize, evaluate, and refine text \citep{flower1981cognitive, du2022understanding}.
Recent efforts in revision research in scientific writing have largely focused on human revision corpora \citep{jiang-etal-2022-arxivedits}, comment-to-edit alignment \citep{darcy-etal-2024-aries}, or paragraph-level automated revision \citep{jourdan-etal-2025-pararev}.

Research focusing on AI aid in scientific writing has, so far, focused on final judgments of quality \citep{Hazra_2026, tang2023evaluating, Hadan2024_dg, Peters2025_hp} and detectability \citep{sarzaeim2023framework, ladha2023ai, flitcroft2024performance}. Character-level edits help to recover revision events, annotate writing processes, predict text quality, and study how writers allocate effort across revision---critical information unavailable in final outputs \citep{conijn2020process, velentzas2024logging, tian2025klicke, miletic2022pro}. Models like IteraTeR \citep{du2022understanding}, arXivEdits \citep{jiang2022arxivedits}, ParaRev \citep{jourdan-etal-2025-pararev}, among others, focus on revisions at sentence-, word-, and paragraph-levels to establish the crucial role of revisions in scientific writing with AI; however, two lingering questions remain.

\begin{figure}[t!]
    \centering
    \includegraphics[trim=20 150 40 25,clip,width=\linewidth]{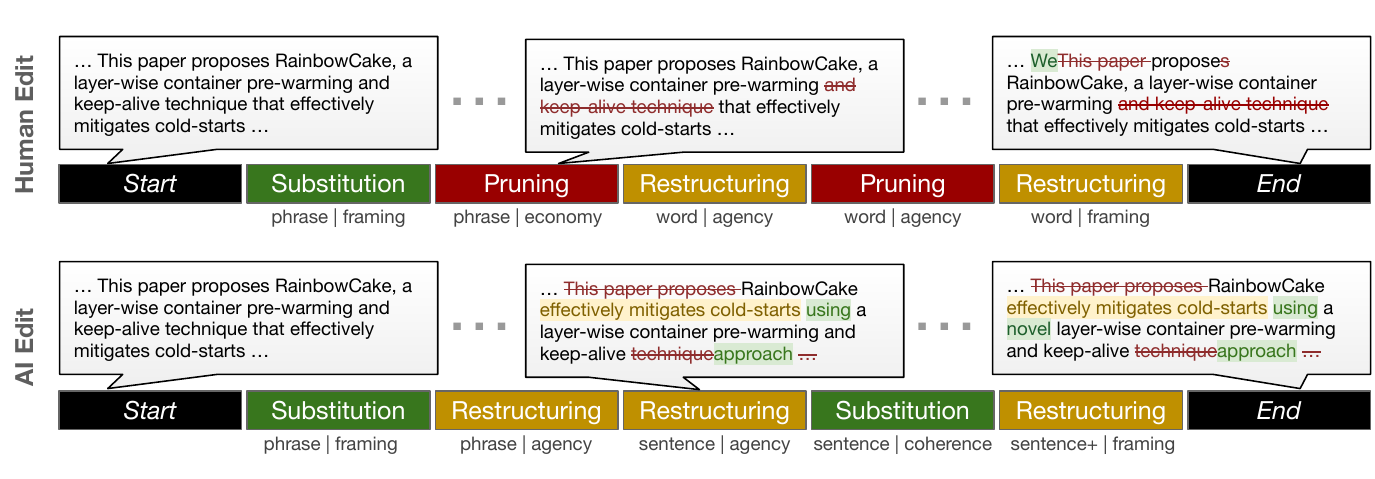}
    \caption{\small \textbf{Ebb and flow of scientific writing.} Human editors capitalize most improvements through substitution and pruning, but eventually fail to alter distributional differences between AI- and human-authored abstracts. LMs offer consistent improvements in sentence-level quality features, but enforcing human behavioral patterns degrades quality. We highlight edit behaviors in the sequence of colored boxes (more on \autoref{sec:revision}). Text below colored boxes indicates the scope of the change, followed by a `$\mid$', followed by the intent of the linguistic change (more on \autoref{sec:ling_prop}).}
    \label{fig:placeholder}
\end{figure}

First, the existing models evaluate either final text quality or local revision behavior, without explicitly relating the two. Second, the revisions are often represented as isolated edit operations or aligned draft pairs. This obscures how draft-level properties drive subsequent edits and how sequences of edits alter a document over time. Instead of evaluating only final outputs or cataloging edits in isolation, \textit{\textbf{we analyze how properties of the current draft predict subsequent revision behavior and how those revisions alter the evolving text by jointly modeling the text state and the revision state.}}


We leverage large-scale experiment data from \cite{Hazra_2026} where participants are segregated into two pools: authors and reviewers. Authors revise an abstract, and reviewers assess if the edited version faithfully reflects the content of the original draft. Importantly, the data corpus captures timestamped keystroke-level edits performed while editing the abstracts. For this study, we infer edit trajectories from these keystroke-level logs by transforming noisy timestamped keyboard and cursor operations into a behaviorally interpretable taxonomy of bursts. Using these metrics, we study fine-grained edit trajectories and analyze distributional differences in revision behavior between human and AI-authored texts. We summarize our contributions as follows: 
\begin{itemize}[leftmargin=*, itemsep=0.1mm]
\item We motivate the joint modeling of text and revision state for scientific writing to formulate human editing strategies and evaluate their efficacy on editing both human-authored and AI-generated abstracts.
\item We observe that human editors uphold local agency ($p<.01$) and structure ($p<.01$) in AI abstracts, but global coherence falls behind ($p<.01$) in human abstracts. Multiple threads of evidence suggest human editors do not close the distributional gap between AI and human-authored abstracts, hinting at the risk of macro-level behavior homogenization.
\item We demonstrate the mixed capabilities of LM editors. They excel locally ($p<.001$), but struggle at the discourse level ($p>.05$); improve the weakest abstracts to average ($p<.001$), but degrade high-quality ones ($p<.05$), indicating LM editors are situationally useful in improving the quality of scientific texts.
\end{itemize}
\section{Data and methods} 
Keystroke-level edits unearth the sequential process of editing.
We act on keystroke logs containing timestamped keyboard and cursor operations performed while editing a piece of scientific text (abstracts) at the character level. Editing behavior emerges from the interplay between a starting text state and a set of revisions that modify it. Our methodological approach involves producing behaviorally meaningful sequences, followed by quantifying the linguistic properties of the texts in sequence.

\subsection{Data}
\begin{wraptable}[19]{R}{0.52\textwidth}
\centering
\small
\vspace{-1.5em}
\begin{tabular}{@{}llr@{}}
\toprule
Group & Variable & Summary \\
\midrule
\multirow[t]{3}{*}{Activity} & Entry pause (s) & 34.00 (308.23) \\
 & Burst duration (s) & 16.00 (108.52) \\
 & Edits per burst & 58.25 (129.23) \\
\multirow[t]{3}{*}{Navigation} & No movement & 95.0 (13.0) \\
 & Backward & 1.4 (6.9) \\
 & Forward & 3.6 (11.0) \\
\multirow[t]{3}{*}{Action} & Insert & 43.9 (41.6) \\
 & Delete & 32.9 (38.5) \\
 & Substitute & 23.2 (31.8) \\
\multirow[t]{3}{*}{Scope} & Word & 9.2 (28.7) \\
 & Phrase & 61.5 (48.0) \\
 & Sentence+ & 29.3 (45.0) \\
\bottomrule
\end{tabular}
\caption{\small Edit burst descriptive statistics for the revision data from \cite{Hazra_2026}. Entry pause and burst duration are reported as \mbox{median} (SD). All remaining variables are reported as mean (SD).
}
\label{tab:data_stats}
\end{wraptable}

We borrow the data from \cite{Hazra_2026}, who use abstract composition as an element of scientific writing. \cite{Hazra_2026} select 45 published papers from notable conference venues in Computer Science. For each paper, they produce 45 respective AI-generated abstracts, controlling for the content and quality of the information.  

The core experiment involves a 2 × 2 between-subject design with the abstract provenance in one arm and its disclosure in the other. Participants with relevant experience are split into author (domain experts) and reviewer (domain experts) pools. Authors are randomly assigned to one of the four treatments. Authors are incentivized to revise the provided abstracts to ensure their edited abstracts are accepted by at least two independent reviewers. Reviewers compare the edited abstracts against the original abstract to provide accept/reject judgments\footnote{For more details about data collection and experiment design, refer to Section 3 in \cite{Hazra_2026}.}. Edits are captured and timestamped at the keystroke level. 

The data thus comprises 45 pairs of abstracts from the selected seed papers: 45 original human abstracts and 45 of their AI-generated counterparts. Each author is randomly assigned to edit three abstracts, generating 891 keystroke-level edit logs. We dropped 22 editing sessions that were deemed AI-assisted. Finally, we were left with 869 keystroke-level edit logs corresponding to 236,033 total edits. The data also contained additional anonymized demographic information about the participants, including education. The descriptive statistics of this data are reported in \autoref{tab:data_stats}.

\subsection{Revision process}
\label{sec:revision}
Analyzing edit trajectories at the character level is noisy and difficult to interpret. Instead, we convert these singular events into behaviorally separable \textbf{\textit{bursts}}. Common practice segments bursts using pause thresholds \citep{Leijten2013_aj}. In our pipeline, we focus on creating bursts that represent contiguous local revisions. Thus, we allow both strong relocation and pauses greater than two seconds to segment the bursts. We take additional measures to detect and merge edge cases, e.g., pausing within a word. In total, we create 3,941 bursts after segmentation\footnote{All p-values reported use FDR (Benjamini-Hochberg) to correct for multiple comparisons.}. 

\paragraph{Burst dimensions.}
We aggregate events within bursts into four dimensions, largely inspired by standard keystroke logging metrics \citep{baaijen_2012_keystroke}.

\begin{itemize}[leftmargin=*, itemsep=0.2mm]
    \item \textbf{\textit{Pauses}} delineate text production vs physical inactivity. We compute two metrics related to time: (1) the average time in seconds prior to starting a burst, and (2) the average pause time within a burst. Both pause metrics are log-transformed due to extreme positive skew. We additionally track (3) the number of edits made within a burst.
    \item \textbf{\textit{Navigation}} tracks relative cursor position as a proxy for attention and grounds the movement of edits. We measure the share of (1) movement among forward, (2) backward, and (3) no movement, quantified as a proportion of the entire text.
    \item \textbf{\textit{Actions}} describe the edit operations: (1) insert, (2) delete, and (3) substitute. We calculate the share among insert, delete, and substitute edits.
    \item \textbf{\textit{Scope}} elaborates the breadth of text necessary for understanding the context of an edit. We categorize scope into (1) word-level, (2) phrase-level, and (3) sentence+ using the pre- and post-burst text difference. 
\end{itemize}

\paragraph{Defining burst behaviors.}
To better interpret the twelve continuous features per burst (three from each of the four burst dimensions above), we cluster the variables with a Gaussian mixture model, choosing up to twenty clusters. The final cluster count is determined by cluster fitness with the Bayesian Information Criterion (BIC) and additional stability analysis. Perfectly collinear features (e.g. action share) are transformed via an isometric log-ratio prior to clustering.

In total, the algorithm chose sixteen clusters. We manually label each according to the centroid averages on the twelve features, naming according to dominant action mix, dominant scope, and salient navigation direction; we focus on observable editing behavior rather than assuming latent states. The edit burst behaviors, described in \autoref{tab:cluster_labels}, fall into four general families with frequencies in parentheses: expansion ($28\%$), pruning ($23\%$), substitution ($30\%$), and restructuring ($19\%$). \textbf{Expansion}, \textbf{pruning}, and \textbf{substitution} relate to edits which primarily add, remove, or replace material, respectively. \textbf{Restructuring} reworks content through multiple edit actions at a broader scope. The cluster fit metrics and feature means per cluster are available in \Cref{tab:app_cluster_means_full,tab:app_gmm_summary,tab:app_gmm_stability} in \autoref{app:methods}.

\begin{table}[t]
\small
\centering
\begin{footnotesize}
\setlength{\tabcolsep}{5pt}
\renewcommand{\arraystretch}{1}
\begin{tabularx}{\linewidth}{L{0.13\linewidth} L{0.25\linewidth} Y}
\toprule
\textbf{Family} & \textbf{Behavior label} & \textbf{Definition} \\
\midrule

Expansion & Phrasal expanding & Phrase scope, insert-only \\
Expansion & Sentence+ expanding & Sentence+ scope, insert-only \\
Expansion & Sentence+ elaborate & Sentence+ scope, insert with minor delete/substitute \\
\midrule

Pruning & Backward word prune & Single-word delete with backward navigate \\
Pruning & Backward span prune & Phrase/sentence scope delete with backward navigate \\
Pruning & Forward word prune & Single-word delete with forward navigate \\
Pruning & Forward span prune & Phrase or sentence scope delete with forward navigate \\
Pruning & Phrasal prune & Phrase scope, deletion-only \\
Pruning & Phrasal condense & Phrase scope delete/substitute \\
\midrule

Substitution & Forward word replace & Single-word replace with forward navigate \\
Substitution & Phrasal replace & Phrase scope, balanced insert/delete \\
Substitution & Phrasal rewrite & Phrase scope, mixed insert/delete/substitute \\
Substitution & Phrasal substitute & Phrase scope, substitute-only \\
Substitution & Phrasal insert+substitute & Phrase scope substitute with added insert \\
\midrule

Restructuring & Sentence+ restructure & Sentence+ scope, mixed insert/delete/substitute  \\
Restructuring & Cross-scope restructure & Mixed-scope, mixed insert/delete/substitute \\

\bottomrule
\end{tabularx}
\end{footnotesize}
\caption{Edit-burst behavior labels. Sentence+ denotes sentence-level or larger editing spans.}
\label{tab:cluster_labels}
\end{table}

\subsection{Linguistic text properties}
\label{sec:ling_prop}

Editing inherently perturbs the linguistic properties of a text. The readability of a scientific text improves when contextual material appears in the topic position and new or salient information is placed in the stress position, where readers expect it \citep{halliday1967notes, gopen1990science}. Clarity, likewise, is enhanced when grammatical subjects mention concrete “characters” and verbs express actions directly, rather than obscuring them through nominalizations \citep{swales2014genre, williams2014style}. We thus need to create an outcome that captures the effect of the behavioral states established in \autoref{sec:revision}. We group 25 computational linguistic metrics, derived from \cite{Hazra_2026}, into five theory-driven dimensions to improve the broader interpretability of results. We focus on descriptive dimensions calculable from the text. Please see \autoref{tab:app_metric_merge} in \autoref{app:methods} for a complete description of all 25 variables.

All dimensions are coded such that higher scores tend to indicate better writing. Each individual metric is standardized in reference to the initial sample collected in \citep{Hazra_2026} for comparability. Thus, the median value for each metric is zero, and a score of one represents a single standard deviation above the average. We take a mean of the individual metrics to compute the final dimension score. The linguistic text properties are defined as follows:
\begin{itemize}[leftmargin=*, itemsep=0.2mm]
\item \textbf{Agency} clarifies the role between the actors (subjects) and their actions (verbs) \citep{Hyland2002_jk, Hyland2005MetadiscourseEI, gopen1990science, halliday2014introduction}. Clear writing makes it easy to identify who is doing what.
\item \textbf{Economy} quantifies how efficiently a text conveys complex information without unnecessary compression \citep{Halliday2004, Biber_Gray_2013}. It represents the tradeoff between information-rich writing and reading difficulty. Hallmarks of poor economy include nominalizations, dense noun clusters, and filler words.
\item \textbf{Structure} describes how the arrangement of words and phrases leads to the primary grammatical relationships \citep{gopen1990science, gibson1998linguistic, williams2014style}. Strong writing prioritizes both succinct links between its core components and balanced composition.
\item \textbf{Coherence} measures how well a text introduces new information and connects threads \citep{Graesser2004_nf, Halliday1976_mk}. Best practice suggests pushing new information later within a sentence, and keeping topics similar across nearby sentences.
\item \textbf{Framing} tracks the placement of context and claims \citep{Swales1990_xr, Hoey2000_qx}. Balanced positioning allows the reader to understand a point and its significance best.
\end{itemize}

\section{Behavioral and linguistic analyses on edit logs}
\label{sec:results}

\begin{table}[t]
  \centering
  \small
  \begin{minipage}{0.48\textwidth}
    \centering
    \begin{tabular}{
        l 
        S[table-format=-1.2, table-space-text-post={$^{**}$}] 
        S[table-format=-1.2, table-space-text-post={$^{**}$}] 
        S[table-format=1.2]
    }
    \toprule
    {$\Delta$ Dimension} & {$\beta_{\text{source}}$} & {$\beta_{\text{pre\_edit}}$} & {$\Delta R^2$} \\
    \midrule
    Agency & -0.05 & -0.34$^{**}$ & 0.10 \\
    Economy & -0.02 & -0.37$^{**}$ & 0.12 \\
    Structure & 0.02 & -0.39$^{**}$ & 0.14 \\
    Coherence & -0.13$^{**}$ & -0.39$^{**}$ & 0.08 \\
    Framing & -0.00 & -0.23$^{**}$ & 0.03 \\
    \bottomrule
    \end{tabular}
    \captionof{table}{\small Effect sizes of text source and pre-edit linguistic properties on the edit outcome at the burst level, and the difference in variance explained between the pre-edit and source. ${*} p < .05$, $^{**} p < .01$.}
    \label{tab:delta_metrics}
  \end{minipage}
  \hfill 
  \begin{minipage}{0.48\textwidth}
    \centering
    \includegraphics[width=\textwidth]{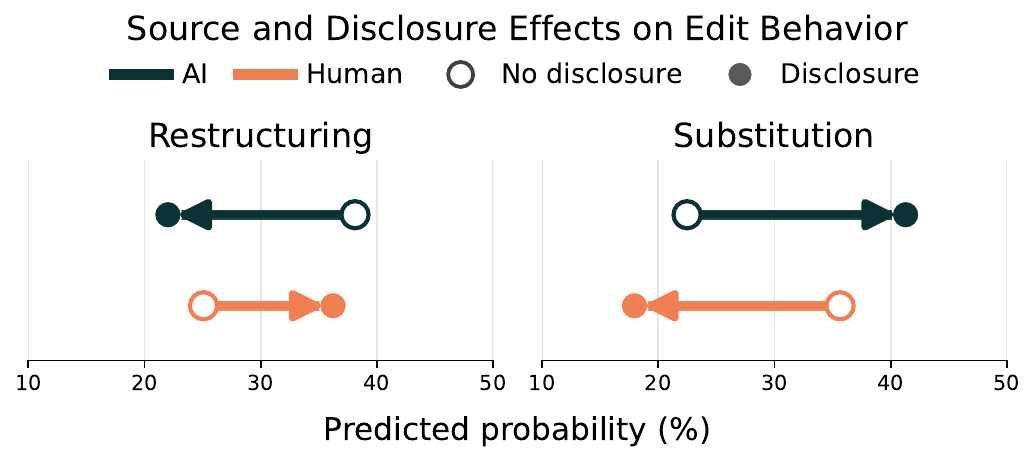}
    \captionsetup{type=figure}
    \vspace{-1.5em}
    \caption{\small Edit behavior changes in those with masters+ education moderated by abstract source (human vs AI) and disclosure (no disclosure vs disclosure)}
    \label{fig:behavior_education}
  \end{minipage}
  
\end{table}

\subsection{Pre-edit linguistic features best explain edit trajectories.}
\label{sec:4.1}
We fit separate linear models predicting the standardized change in each of the linguistic properties from abstract provenance (AI vs. Human) and the corresponding standardized pre-edit score. We additionally controlled for the burst position within the session, and the source paper of the abstract. In all five metrics, pre-edit linguistic properties substantially outweigh provenance in predictive power. Lower initial scores were associated with larger improvements, evidenced by the negative pre-edit metric effect sizes ($\beta_{\text{pre\_edit}}$) ranging from $-0.23$ to $-0.39$ (all $p < .01$) shown in \autoref{tab:delta_metrics}. In contrast, provenance explained relatively little variance; only coherence showed a small significant effect ($\beta_{\text{source}} = -0.13, p<.01$). Even so, the abstract source explained only a small portion of unique variance compared to pre-edit metrics ($R^2_\text{source} = 0.004, R^2_\text{pre\_edit} = 0.085$)\footnote{Additional details on the partial $R^2$ are shown in \autoref{tab:delta_metric_full} in \autoref{app:results}.}.

\subsection{Education moderates behavior related to AI disclosure.}
Multinomial regression accounting for education level reveals several significant differences in editing behavior. Those with bachelors or lower use the four burst categories with relative similarity for AI and human authored abstracts. However those with masters or higher education ($37\%$ of the data) vary behaviorally depending on both source and disclosure. \autoref{fig:behavior_education} shows that for AI abstracts, disclosure is associated with severe drops in restructuring ($\Delta=-16.1\%, p<.001$) in favor of substitution ($\Delta=+18.8\%, p<.001$). Disclosing human source had the opposite effect: restructuring ($\Delta=+11.1\%, p<.05$) and substitution ($\Delta=-17.7\%, p<.01$). This reflects the common stigma against AI text. Editors are only willing to make restructuring changes to AI text without disclosure, and much rather perform substitutions when they know the text is AI. The full statistics for all education levels and contrasts are available in \autoref{tab:disclosure_behavior_appendix}.

\subsection{Human revision does not improve final abstract quality.}
\label{sec:4.2}
\begin{figure}[t]
  \centering
  
  \begin{subfigure}{0.48\textwidth}
    \centering
    \includegraphics[width=\textwidth]{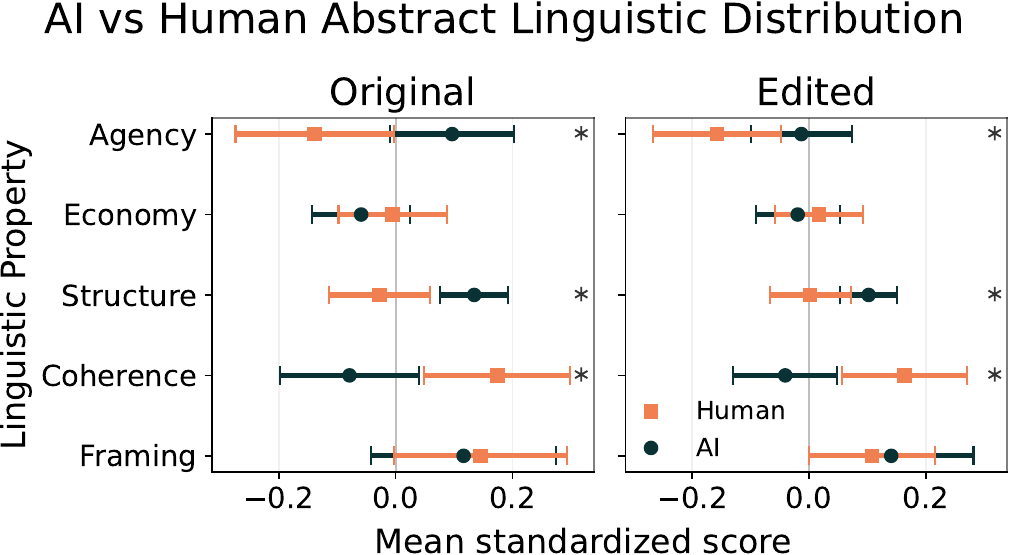}
    \caption{}
    \label{fig:basic_stats}
  \end{subfigure}
  \hfill 
  \begin{subfigure}{0.48\textwidth}
    \centering
    \includegraphics[width=\textwidth]{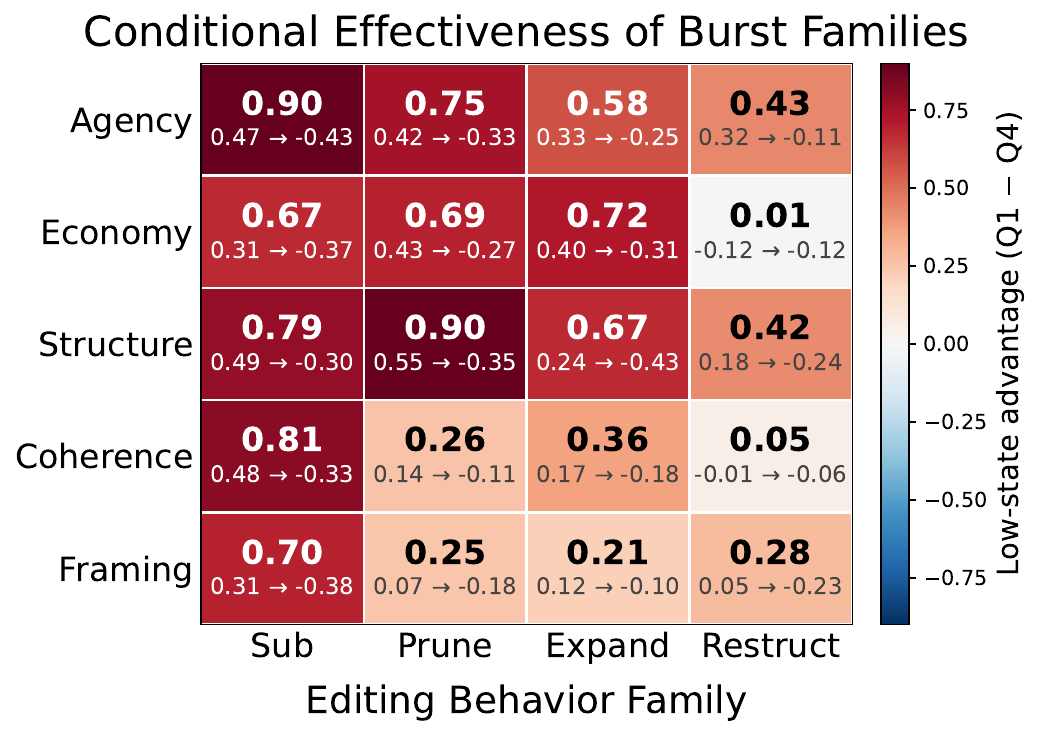}
    \caption{}
    \label{fig:behavior_efficacy}
  \end{subfigure}
  \caption{(a) Linguistic properties of original and edited AI-and-human authored abstracts. (b) Editing behavior efficacy in low and high scoring regions of the abstracts. For the top value, higher means more beneficial in weaker states. The bottom values are the mean Q1 and Q4 respectively.}
\end{figure}
\textbf{Human and AI abstract linguistic distributions remain divergent.}
\cite{Hazra_2026} reports significant differences in number of edits across source and disclosure, which suggests certain linguistic improvements may be sensitive to the treatment condition. Because revision is a consequence of the initial properties of the abstract, we first conduct a paired t-test between human-AI abstract counterparts, visualized in \autoref{fig:basic_stats}. AI abstracts were significantly stronger than human abstracts in sentence-level agency ($d_z = 0.45, p=.008$) and structure ($d_z = 0.54, p=.004$), but underperformed in global coherence ($d_z = -0.47, p=.007$). The only two significant changes post-editing occurred in AI, where Agency decreased ($d_z = -0.60, p=.001$) and Economy increased ($d_z = 0.46, p=.009$). However, the original distributional differences between human and AI authored abstracts remain, despite an average of 300 individual edit actions, and 6 bursts per abstract editing session. Ultimately, human editors fail to alter the distributional differences between AI- and human-authored abstracts. Please refer to \Cref{tab:initial_states,tab:post_states,tab:app_delta_abs} in \autoref{app:results} for the full statistics.

\textbf{Pre-edit linguistic features moderate behavior efficacy.}
We further investigate the result from \autoref{sec:4.1}. Pre-edit regions with lower scores show the strongest improvement. We calculate the mean improvement in the highest (Q4) and lowest (Q1) quartiles, attributed to their respective burst behavior family. \autoref{fig:behavior_efficacy} shows that burst behaviors vary in efficacy, moderated by the linguistic property and its starting state. Substitution most strongly and broadly repairs weak states, suggesting that replacing text in place is the strongest method to improve low-quality starting states. Pruning is selectively strong for agency, economy, and structure. Intuitively, deleting text should help de-clutter and improve the structure of bloated text. Adding new text is similarly helpful for sentence-level metrics, but not as strong as pruning; likely because expansion without pruning means the original text dilutes the performance improvements. Restructuring shows a relatively weak effect in repairing low-quality regions of the abstract. Finally, all of the Q4 behavior efficacies are negative, suggesting that humans tend to hinder the highest quality abstracts. The full metrics and confidence intervals are reported in \autoref{tab:behavior_ling_q1_q4} in \autoref{app:results}. 

\begin{takeawaybox}
\small
\textbf{Implications:} Human editors capitalize on the lower quality regions of abstracts, with diminishing returns on regions of higher quality. They tend to make the most improvements through substitution and pruning, rather than heavier restructuring. While those with a masters' degree or above are willing to restructure AI text, disclosure is associated with a reversed effect, inducing significantly more substitution. Ultimately, human editors fail to alter the distributional differences between AI- and human-authored abstracts.

\end{takeawaybox}
\section{Language models as editing assistants}
\label{sec:lm_exp}
Poor human editing outcomes seen in \autoref{sec:4.2} suggest ample opportunity for the growing adoption of LMs as assistants \citep{liang2025widespread}. We test two common applications of AI for scientific abstracts to identify whether LMs can enhance or compensate for deficiencies in human editing. First, revising an already written abstract (zero-shot editing), and second, revising a specific component of the abstract (scoped assistant). We experiment with the latest versions of GPT (5.4) and Claude (Opus 4.6), reflecting upon the systems people are most likely to use for similar tasks in the current time. For further breakdown of model specific results, please see \autoref{app:lm_exp_diff}.

\subsection{Use case 1: LM as a zero-shot editor}
We prompt the LM to edit the 45 abstract pairs to improve their quality using two separate task descriptions.
\begin{itemize}[leftmargin=*, itemsep=0.2mm]
\item \textbf{Unguided} provides general instructions about scientific abstracts (the same instructions seen by participants in \citet{Hazra_2026}).~\textbf{Unguided} represents how a user might interact with an LM naturally.
\item \textbf{Rubric} provides a rubric capturing all linguistic properties described in \autoref{sec:ling_prop} (see \autoref{tab:app_metric_merge} in \autoref{app:methods}). \textbf{Rubric} grounds the LM in metrics that make an abstract stronger. Detailed prompts can be found in \autoref{app:prompts}

\end{itemize}

\paragraph{LMs improve locally but struggle globally.} As established in \autoref{sec:4.2}, humans do not improve the linguistic dimensions significantly. In contrast, \textbf{Unguided} shows significant improvements in economy, further improved by LMs with \textbf{Rubrics} ($d_z=0.53, p<.001$). \textbf{Rubrics} additionally improve agency ($d_z=0.50, p<.001$) and structure ($d_z=0.37, p<.001$) during editing. While framing is significant in \textbf{Unguided} generation, the differences remain underpowered when a consistent \textbf{Rubric} is provided. Congruent with the results from \autoref{sec:4.2}, LMs as editors largely improve on the same distributions that were originally strong in AI-authored abstracts. Overall, the results imply: while LMs robustly improve sentence-level metrics, they struggle at the discourse level. Complete statistical test results are available in \Cref{tab:app_lm_zeroshot,tab:app_lm_zeroshot_comp} in \autoref{app:lm_exp}.

\begin{figure}[t]
  \centering
  \begin{subfigure}{0.45\textwidth}
    \centering
    \includegraphics[width=\textwidth]{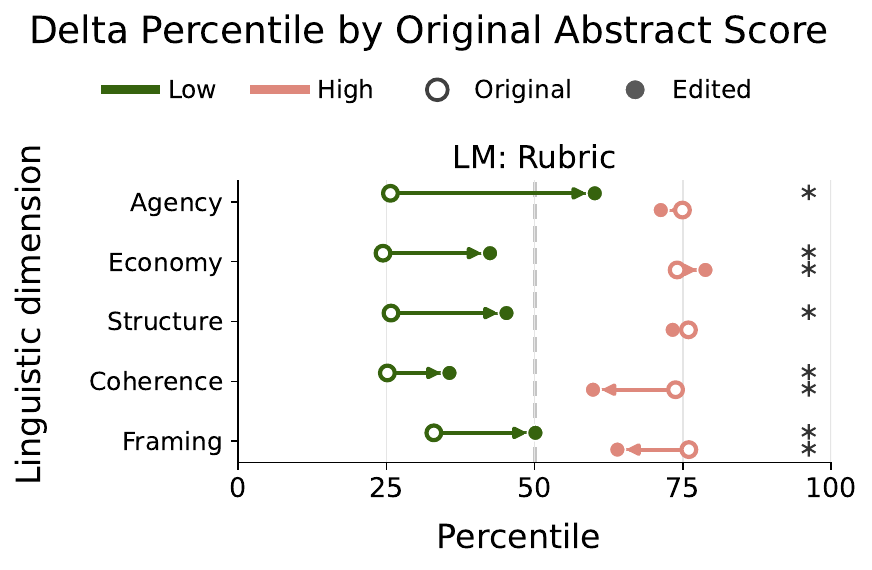}
    \caption{}
    \label{fig:lm_prepost_percentiles}
  \end{subfigure}
  \hfill 
  \begin{subfigure}{0.45\textwidth}
    \centering
    \includegraphics[width=\textwidth]{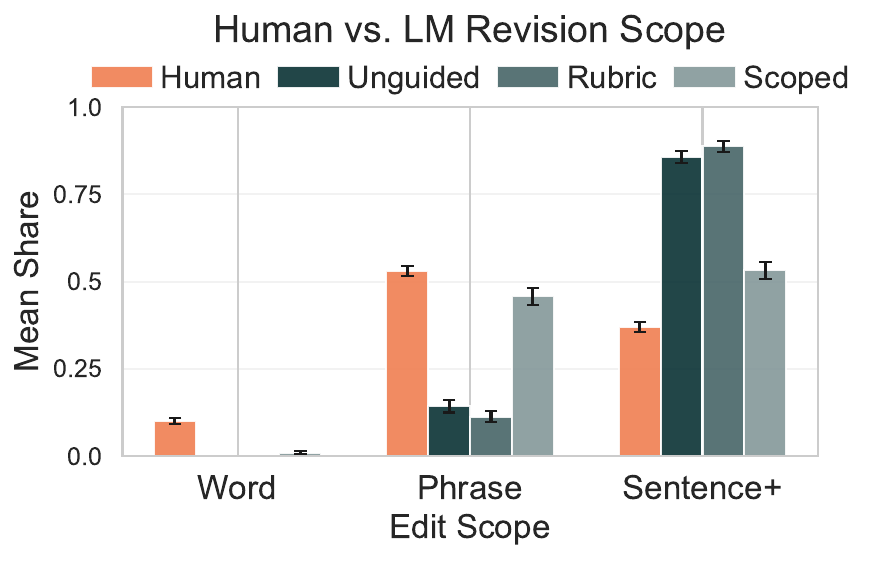}
    \caption{}
    \label{fig:scope_alignment}
  \end{subfigure}
  \caption{(a) LMs fail to improve low-scoring abstracts to above average. Besides economy, high-scoring abstracts suffer or stay the same. (b) LMs consistently performed larger-scoped edits than humans. Prompting with the desired scope improves alignment, but still produces significantly larger edits.}
\end{figure}

\paragraph{LM editors make bad abstracts average, and good abstracts worse.} \autoref{fig:lm_prepost_percentiles} delineates linguistic improvement of the bottom and top-scoring abstracts. For this analysis, abstracts are split by median based on each linguistic dimension. The lowest-scoring abstracts consistently receive the strongest improvements, lifting them from an average percentile of $26.8\%$ to $46.7\%$. All individual metric gains tested significant and are available in \autoref{tab:app_delta_zeroshot_percentile} in \autoref{app:lm_exp}. On the other hand, the highest-scoring abstracts often had no significant gain. Rather, coherence dropped modestly ($13.9\%, p<.001$). Economy showed signs of improvement ($4.8\%, p<.05$). We find LM editors may enhance the lowest-scoring abstracts up to the mean; they often struggle to maintain consistency and improve the best abstracts, often actively hindering them.

\subsection{Use case 2: LM as a scoped assistant}
Humans edit in targeted, interpretable bursts. Aligned assistants should edit in a way that is congruent with human intent. We measure LM editing alignment in this case by asking the LMs to improve the abstract scoped to the same chunk of text a human edited. Each human editing burst could correspond to a single phrase, sentence, or the entire abstract. We sample 200 such bursts from the original 90 abstracts using inverse distribution weighting to ensure a diverse set of editing actions, scope, and outcomes. We utilize the same \textbf{Unguided} and \textbf{Rubric} experiment descriptions as in the zero-shot setup. In addition, we add another description, \textbf{Scoped}, that mandates explicitly that the LM must adhere to the exact editing scope of the human (word, phrase, sentence+).

\paragraph{LMs betray editing scope.} \autoref{fig:scope_alignment} shows that even when conditioned on the same scope, models often ignore the scope and make wider-reaching changes across sentences. This misalignment is difficult for a human to verify visually, especially with increased text lengths, and can compound over an editing session. \textbf{Scoped} results in better alignment with human editing intentions, but still results in significantly more sentence+ revisions. All statistically significant differences are reported in \autoref{tab:app_scope_comparison} in \autoref{app:lm_exp}.

\paragraph{LMs with scoped edits are weaker at improving the worst abstracts.} \textbf{Unguided} and \textbf{Rubric} LM editors at the burst level follow the same trends as in the zero-shot setup. We point readers to \autoref{tab:app_burst_prepost} in \autoref{app:lm_exp} for additional details. More interestingly, \textbf{Scoped} editors underperform compared to \textbf{Rubric} in the lowest scoring abstracts (average $\Delta = -4.6\%$, all individual tests $p<.05$, besides coherence $p=.351$). Contrastingly, for high-scoring abstracts, their scores are not significantly different. This suggests that LMs do not edit their strongest when forced to align with human scope. Further post-training efforts should focus on aligning LMs to faithfully follow human editing intentions. All statistics are reported in \autoref{tab:rubric_vs_scoped} in \autoref{app:lm_exp}. 

\begin{takeawaybox}
\small
\textbf{Implications:} LMs offer consistent improvements on sentence-level features, but only by bringing the lowest-scoring abstracts to average. Language models also consistently make larger-scoped edits than humans. Forcing LMs to edit closer to human scopes harms performance on the lowest-scoring abstracts.
\end{takeawaybox}
\section{Discussion}

\paragraph{Stigma and AI writing.}

Nature quotes, "Tools such as ChatGPT threaten transparent science" \citep{Nature2023ChatGPTGroundRules}. Is this stance generalizable? We find that education is a key marker of stigma in responses to AI-assisted scientific writing, specifically among individuals with a master's degree or above. However, stigma towards AI can encompass a wide range of behaviors: intent to avoid AI detection \citep{fang2025shapeswritersdecisionsdisclose}, distrust in AI \citep{10.1145/3742413.3789076}, and even lower perceived ownership \citep{draxler2024ai}. Discovering more complex links between perceptions of AI and writing behavior is a fruitful direction to explore in future work. But it remains clear that the usefulness of AI for scientific writing hinges on the willingness to adopt and meaningfully engage with it. While we see efficiency gains from using AI assistance in revising bad-to-moderate scientific texts, this resistance may be concentrated and percolated among those with stronger disciplinary writing conventions. 

\paragraph{Homogenizing behavior.}

LMs tend to reduce self-correction and increase irrational decision-making \citep{cheng_sycophancy, sharma2024towards, bhattacharya2026good}. Our results imply parallels. Human editors focused on the most visible weaknesses in a text, yet these interventions rarely closed the broader quality gap between AI- and human-generated abstracts. This points to an alarming situation in which homogenized AI-generated text may prevail even after human editing. Cognitively, repeated exposure to apparently fluent but suboptimally structured AI-generated scientific texts may explain the tendency towards superficial edits \citep{abdulhai2026llms}. LM editors are in no position to rescue; in fact, they can often make the situation worse by degrading the quality of high-quality abstracts. Our findings mirror findings from \citep{noy2023experimental,doshi2024generative,agarwal2025ai} where larger gains with LM assistance are observed with weaker writers or weaker abstracts, but to raise the quality to a stylistically homogeneous baseline.
\section{Additional Related work}
\paragraph{AI writing and edit-based evaluation.}
The spread of AI writing tools makes it essential to review how humans interact with, assess, and edit AI-generated texts. AI-generated creative texts exhibit 3x to 10x more stylistic weaknesses than human-authored texts \citep{chakrabarty2024art}. Editing human writing and AI-generated text also presents challenges, as \cite{chakrabarty2025can} observes that although edits for AI-generated texts are broadly similar, they were needed much more extensively compared to human-authored texts. IteraTeR \citep{du2022understanding}, arXivEdits \citep{jiang-etal-2022-arxivedits}, ParaRev \citep{jourdan-etal-2025-pararev} contribute to scientific revision being an iterative and structured process, with authors making systematic edits for style, precision, simplification, grammar, and formatting across full-paper versions rather than merely altering final text superficially. ScholaWrite \citep{le2025scholawrite} extends this line of research by providing keystroke-level traces of end-to-end scholarly writing, enabling the study of revision at a finer temporal resolution. 

\paragraph{AI for scientific writing.}

LMs are increasingly shaping scientific writing, from assistant plugins like Writefull\footnote{\url{https://www.writefull.com/}} being integrated into the editing platforms of Overleaf to tools that support drafting, revision, and even full paper generation \citep{Ito2019DiamondsIT, Bezerra2021NEWRITERAT,
wang-etal-2019-paperrobot, kanna2024much, Liebling2025TowardsAA, Kousha2025HowMA, lu2024ai}. At the same time, anecdotal cases of seemingly AI-generated papers and peer reviews, along with broader critiques of generative AI in academia, have raised concerns about authenticity and quality \citep{cybernews_academic_2023, anthes_ai_2023, oransky_papers_2024, liang2024mapping}. Corpus-level research further suggests that LLM-generated content is becoming widespread in scientific writing and may be contributing to more standardized or declining prose quality across disciplines \citep{matsui2024delving,lemire2024will,geng2024chatgpt,kobak2025delving}. More recently, \cite{Hazra_2026} foregrounds the need to understand how humans interact with AI-generated texts, how much agency they retain, how edited AI-generated texts are perceived by an external evaluator, and how these systems reshape scientific communication norms \citep{reza2025co}.

\section{Conclusion}
LMs are reshaping writing and editing, but little is known about how authors revise AI-generated scientific text. We push the community towards a process revision mindset, using bursts in edit trajectories to reveal patterns in editing strategy. We highlight heterogeneous editing patterns by human editors at different scopes (word to sentence), but it is often ineffective to change an AI-generated text extensively, suggesting broader implications of behavior homogenization. While LMs tend to improve sentence-level features during editing, the benefits remain concentrated on the lowest-scoring abstracts, telling a cautionary tale of LM assistance in scientific writing. 

\section{Limitations}

\textbf{Limited sample.} The abstracts inherited from \citep{Hazra_2026} are limited to the domain of computer science. Other domains, such as math, physics, or the humanities, may invite differing editing behaviors due to differing underlying text features. Additionally, abstracts are only a small portion of all scientific writing. Understanding how humans revise longer-form scientific writing remains an interesting avenue for future work. However, our analytical framework generalizes to other domains and forms of scientific writing.

\textbf{Expert editors.} We see more interesting results with the editors who are best aligned with the task. Our current sample hosts a rather small pool of experts, and an in-depth exploration with an expert and senior author can be insightful. 

\section*{Acknowledgments}

This material is based upon work supported by the Ai2 Faculty
Research Award and the OSU Kirwan Institute. We are also grateful
to Ohio Supercomputing Center (OSC) and NAIRR Pilot Program (NAIRR250264) for providing the compute resources for this work.
\newpage

\bibliography{colm2026_conference}
\bibliographystyle{colm2026_conference}

\appendix
\newpage
\section{Reproducibility statement}
We plan to release the full analysis code, data, and preprocessing scripts upon publication.

\section{Language model usage disclosure}
We leveraged AutoDiscovery \citep{agarwal2025autodiscovery}, an LM-based data-driven discovery tool, to perform an initial exploration of our data. While we did not use any specific result from the output, it generally informed our analytic directions. We use LMs (e.g., Grammarly) to check grammatical and structural mistakes. 
\section{Additional details on methods}
\label{app:methods}

Additional results relating to \Cref{tab:app_cluster_means_full,tab:app_gmm_summary,tab:app_gmm_stability,tab:app_metric_merge}.


\begin{sidewaystable}[p]
\centering
\scriptsize
\resizebox{\textheight}{!}{%
\begin{tabular}{lcccccccccccc}
\toprule
Cluster & Log pause & Entry pause & Edits & No Nav & Back & Fwd & Insert & Delete & Sub & Word & Phrase & Sent+ \\
\midrule
Phrasal expanding & 5.918 & 120.189 & 45.489 & 0.986 & 0.003 & 0.011 & 1.000 & 0.000 & 0.000 & 0.000 & 1.000 & 0.000 \\
Phrasal substitute & 6.084 & 104.342 & 16.399 & 0.982 & 0.005 & 0.013 & 0.000 & 0.000 & 1.000 & 0.000 & 1.000 & 0.000 \\
Phrasal prune & 6.340 & 99.230 & 10.572 & 0.953 & 0.014 & 0.033 & 0.000 & 1.000 & 0.000 & 0.000 & 1.000 & 0.000 \\
Phrasal replace & 6.362 & 121.257 & 21.733 & 0.972 & 0.013 & 0.015 & 0.507 & 0.493 & 0.000 & 0.000 & 1.000 & 0.000 \\
Sentence+ restructure & 6.281 & 109.023 & 106.813 & 0.958 & 0.015 & 0.027 & 0.274 & 0.439 & 0.287 & 0.000 & 0.000 & 1.000 \\
Forward word replace & 3.784 & 165.563 & 4.901 & 0.884 & 0.000 & 0.116 & 0.370 & 0.230 & 0.401 & 1.000 & 0.000 & 0.000 \\
Sentence+ expanding & 5.978 & 142.310 & 203.030 & 0.992 & 0.000 & 0.008 & 1.000 & 0.000 & 0.000 & 0.000 & 0.000 & 1.000 \\
Forward word prune & 0.000 & 104.514 & 1.000 & 0.845 & 0.000 & 0.155 & 0.000 & 1.000 & 0.000 & 1.000 & 0.000 & 0.000 \\
Cross-scope restructure & 7.180 & 65.715 & 40.402 & 0.869 & 0.057 & 0.074 & 0.379 & 0.397 & 0.224 & 0.129 & 0.381 & 0.490 \\
Backward span prune & 0.000 & 185.090 & 1.000 & 0.762 & 0.238 & 0.000 & 0.282 & 0.576 & 0.142 & 0.000 & 0.475 & 0.525 \\
Phrasal rewrite & 5.972 & 117.693 & 53.936 & 0.992 & 0.002 & 0.006 & 0.319 & 0.331 & 0.351 & 0.000 & 1.000 & 0.000 \\
Sentence+ elaborate & 6.419 & 120.703 & 125.610 & 0.949 & 0.025 & 0.026 & 0.632 & 0.177 & 0.191 & 0.000 & 0.197 & 0.803 \\
Forward span prune & 0.000 & 195.389 & 1.000 & 0.600 & 0.000 & 0.400 & 0.209 & 0.663 & 0.128 & 0.000 & 0.452 & 0.548 \\
Phrasal insert+substitute & 6.141 & 74.764 & 38.512 & 0.984 & 0.006 & 0.010 & 0.498 & 0.000 & 0.502 & 0.000 & 1.000 & 0.000 \\
Backward word prune & 0.963 & 78.386 & 1.353 & 0.768 & 0.232 & 0.000 & 0.132 & 0.794 & 0.074 & 1.000 & 0.000 & 0.000 \\
Phrasal condense & 6.082 & 140.812 & 22.000 & 0.984 & 0.005 & 0.011 & 0.000 & 0.497 & 0.503 & 0.000 & 1.000 & 0.000 \\
\bottomrule
\end{tabular}
}
\caption{Cluster means across pause/activity, navigation, action, and scope features.}
\label{tab:app_cluster_means_full}
\end{sidewaystable}

\begin{table}
\centering
\begin{small}
\begin{tabular}{lcccccc}
\toprule
$k$ & Covariance & BIC & Mean max posterior & Min cluster $n$ & Smallest cluster share & Converged \\
\midrule
2 & full & 22779.49 & 1.0000 & 1572 & 0.3989 & True \\
3 & full & 8908.56 & 0.9991 & 620 & 0.1573 & True \\
4 & full & -2338.91 & 0.9998 & 275 & 0.0698 & True \\
5 & full & -8593.44 & 0.9983 & 275 & 0.0698 & True \\
6 & full & -9932.09 & 0.9955 & 317 & 0.0804 & True \\
7 & full & -18591.95 & 0.9987 & 119 & 0.0302 & True \\
8 & full & -20675.32 & 0.9981 & 144 & 0.0365 & True \\
9 & full & -23794.34 & 0.9980 & 94 & 0.0239 & True \\
10 & full & -27518.06 & 0.9982 & 94 & 0.0239 & True \\
11 & full & -30685.60 & 0.9991 & 35 & 0.0089 & True \\
12 & full & -31232.85 & 0.9947 & 73 & 0.0185 & True \\
13 & full & -34489.82 & 0.9985 & 75 & 0.0190 & True \\
14 & full & -36391.13 & 0.9984 & 35 & 0.0089 & True \\
15 & full & -35873.42 & 0.9952 & 15 & 0.0038 & True \\
\textbf{16} & \textbf{full} & \textbf{-43035.56} & \textbf{0.9995} & \textbf{40} & \textbf{0.0101} & \textbf{True} \\
17 & full & -40712.56 & 0.9974 & 14 & 0.0036 & True \\
18 & full & -40137.37 & 0.9960 & 13 & 0.0033 & True \\
19 & full & -42742.12 & 0.9884 & 11 & 0.0028 & True \\
\bottomrule
\end{tabular}
\end{small}
\caption{Gaussian mixture model selection summary. Cluster size k=16 chosen.}
\label{tab:app_gmm_summary}
\end{table}
\begin{table}
\centering
\begin{tabular}{lccc}
\toprule
Statistic & ARI vs. reference & Mean max posterior & Min cluster $n$ \\
\midrule
count & 25.0000 & 25.0000 & 25.0000 \\
mean & 0.8720 & 0.9960 & 19.4800 \\
std & 0.0580 & 0.0030 & 8.5010 \\
min & 0.7450 & 0.9890 & 8.0000 \\
25
50
75
max & 0.9450 & 1.0000 & 37.0000 \\
\bottomrule
\end{tabular}
\caption{Gaussian mixture model stability summary across repetitions with chosen cluster size 16.}
\label{tab:app_gmm_stability}
\end{table}
\begin{table}[t] 
\begin{center}
\begin{small} 
\begin{tabular}{p{0.12\textwidth} p{0.26\textwidth} p{0.55\textwidth}}
\toprule
\textbf{Dimension} & \textbf{Metric} & \textbf{Meaning} \\
\midrule

Agency & lexical-verb ratio & Proportion of verbs that carry semantic weight to all finite verbs. \\
Agency & passive-voice share & The number of sentences that use passive constructions. \\
Agency & character-as-subject & How often grammatical subject of a sentence refers to a real actor or ``character.'' \\
Agency & agentless-passive & passive clauses omitting a by-phrase agent. \\
Agency & metadiscourse we-density & explicit use of first-person pronouns (\textit{I}/\textit{We}) in sentences that introduce claims. \\
\midrule

Economy & nominalization density & Measures how often verbs and adjectives are turned into abstract nouns. \\
Economy & abstract-congestion & Detects “congested” sentences where an abstract-noun subject is accompanied by additional nominalizations. \\
Economy & compound-noun rate & Captures the number of long noun strings (three or more consecutive nouns). \\
Economy & prep. phrase density & Counts \textit{of}/\textit{in}/\textit{for} relations per 100 words. \\
Economy & filler-word density & Counts filler words like \textit{very}, \textit{actually}, \textit{basically}. \\
\midrule

Structure & subject-onset distance & Measures how many tokens occur before the main grammatical subject. \\
Structure & initial-momentum & How soon the main verb appears relative to the sentence start. \\
Structure & subject-verb gap & The number of words between a subject and its verb. \\
Structure & verb-object gap & Tokens separate a verb from its direct object. \\
Structure & post-verb density & Count of subordinate clauses or heavy additions that follow the main verb. \\
\midrule

Coherence & cohesion-chain strength & How well consecutive sentences connect via repeated or echoed content words. \\
Coherence & old-to-new ordering & The relative position of previously unseen words in each sentence. \\
Coherence & topic-continuity & Whether successive sentences share the same grammatical subject. \\
Coherence & stress-position rate & How often the stress position is occupied by a content word. \\
\midrule

Framing & pivot-explicitness & Presence of adversative markers (\textit{but}, \textit{however}, \textit{yet}) indicating a shift. \\
Framing & condition-and-cost & Whether problem statements articulate both the condition and the consequence. \\
Framing & claim placement & Position where the main contribution or result (e.g., ``we show'') appears. \\

\bottomrule
\end{tabular}
\end{small}
\end{center}
\caption{Text property dimensions and metric definitions. Agency, Economy, and Structure can be interpreted at a sentence level, whereas Coherence and Framing are interpretable only across the broader text. Metrics definitions are derived from \citep{Hazra_2026}. Return to \autoref{sec:ling_prop}.}\label{tab:app_metric_merge}
\end{table}

\section{Additional details on results}
\label{app:results}

Additional results relating to \Cref{tab:delta_metric_full,tab:disclosure_behavior_appendix,tab:behavior_ling_q1_q4,tab:initial_states,tab:post_states,tab:app_delta_abs}.

\begin{table}
\centering
\begin{tabular}{lccccccc}
\toprule
{$\Delta$ Dimension} & $\beta_{\text{source}}$ & $p_{\text{source}}$ & {$R^2_{\text{source}}$} & {$\beta_{\text{pre\_edit}}$} & {$p_{\text{pre\_edit}}$} & {$R^2_{\text{pre\_edit}}$} & {$R^2_\text{full}$} \\
\midrule
Agency & -0.0463 & 0.1493 & 0.0006 & -0.3362 & 0.0000 & 0.0997 & 0.1170 \\
Economy & -0.0156 & 0.6211 & 0.0001 & -0.3660 & 0.0000 & 0.1154 & 0.1275 \\
Structure & 0.0210 & 0.5109 & 0.0001 & -0.3912 & 0.0000 & 0.1395 & 0.1530 \\
Coherence & -0.1268 & 0.0001 & 0.0040 & -0.3902 & 0.0000 & 0.0849 & 0.0959 \\
Framing & -0.0002 & 0.9944 & 0.0000 & -0.2286 & 0.0000 & 0.0310 & 0.0422 \\
\bottomrule
\end{tabular}
\caption{Full results from \autoref{tab:delta_metrics}, including partial $R^2$, the $R^2$ from the entire model, and the p-values.}
\label{tab:delta_metric_full}
\end{table}
\begin{table}
\begin{small}
\begin{tabular}{lllccccc}
\toprule
Education & Contrast & Behavior & Con. left & Con. right & Difference & {$p_\text{FDR}$} \\
\midrule
College & AI-withInfo - AI-noInfo & Expansion & 0.202 & 0.193 & +0.009 & 0.9540 \\
College & AI-withInfo - AI-noInfo & Pruning & 0.161 & 0.244 & -0.083 & 0.0329 \\
College & AI-withInfo - AI-noInfo & Restructuring & 0.305 & 0.291 & +0.014 & 0.9540 \\
College & AI-withInfo - AI-noInfo & Substitution & 0.332 & 0.272 & +0.060 & 0.4709 \\
College & H-withInfo - H-noInfo & Expansion & 0.251 & 0.254 & -0.003 & 0.9748 \\
College & H-withInfo - H-noInfo & Pruning & 0.170 & 0.198 & -0.027 & 0.7171 \\
College & H-withInfo - H-noInfo & Restructuring & 0.306 & 0.295 & +0.010 & 0.9540 \\
College & H-withInfo - H-noInfo & Substitution & 0.273 & 0.253 & +0.020 & 0.9540 \\
College & AI-noInfo - H-noInfo & Expansion & 0.193 & 0.254 & -0.061 & 0.3757 \\
College & AI-noInfo - H-noInfo & Pruning & 0.244 & 0.198 & +0.046 & 0.4709 \\
College & AI-noInfo - H-noInfo & Restructuring & 0.291 & 0.295 & -0.004 & 0.9748 \\
College & AI-noInfo - H-noInfo & Substitution & 0.272 & 0.253 & +0.019 & 0.9540 \\
College & AI-withInfo - H-withInfo & Expansion & 0.202 & 0.251 & -0.049 & 0.4709 \\
College & AI-withInfo - H-withInfo & Pruning & 0.161 & 0.170 & -0.009 & 0.9540 \\
College & AI-withInfo - H-withInfo & Restructuring & 0.305 & 0.306 & -0.001 & 0.9816 \\
College & AI-withInfo - H-withInfo & Substitution & 0.332 & 0.273 & +0.059 & 0.4709 \\
\midrule
Masters+ & AI-withInfo - AI-noInfo & Expansion & 0.190 & 0.248 & -0.059 & 0.1169 \\
Masters+ & AI-withInfo - AI-noInfo & Pruning & 0.177 & 0.146 & +0.032 & 0.2928 \\
Masters+ & AI-withInfo - AI-noInfo & Restructuring & 0.220 & 0.381 & -0.161 & 0.0004 \\
Masters+ & AI-withInfo - AI-noInfo & Substitution & 0.413 & 0.225 & +0.188 & 0.0005 \\
Masters+ & H-withInfo - H-noInfo & Expansion & 0.269 & 0.175 & +0.094 & 0.0359 \\
Masters+ & H-withInfo - H-noInfo & Pruning & 0.189 & 0.218 & -0.029 & 0.4170 \\
Masters+ & H-withInfo - H-noInfo & Restructuring & 0.362 & 0.251 & +0.111 & 0.0377 \\
Masters+ & H-withInfo - H-noInfo & Substitution & 0.180 & 0.356 & -0.177 & 0.0034 \\
Masters+ & AI-noInfo - H-noInfo & Expansion & 0.248 & 0.175 & +0.073 & 0.0724 \\
Masters+ & AI-noInfo - H-noInfo & Pruning & 0.146 & 0.218 & -0.072 & 0.0335 \\
Masters+ & AI-noInfo - H-noInfo & Restructuring & 0.381 & 0.251 & +0.130 & 0.0115 \\
Masters+ & AI-noInfo - H-noInfo & Substitution & 0.225 & 0.356 & -0.132 & 0.0197 \\
Masters+ & AI-withInfo - H-withInfo & Expansion & 0.190 & 0.269 & -0.080 & 0.0514 \\
Masters+ & AI-withInfo - H-withInfo & Pruning & 0.177 & 0.189 & -0.011 & 0.7159 \\
Masters+ & AI-withInfo - H-withInfo & Restructuring & 0.220 & 0.362 & -0.142 & 0.0038 \\
Masters+ & AI-withInfo - H-withInfo & Substitution & 0.413 & 0.180 & +0.233 & 0.0001 \\
\bottomrule
\end{tabular}
\end{small}
\caption{Disclosure effects on burst behavior probabilities. The full results corresponding to \autoref{fig:behavior_education}. "H" corresponds to human.}
\label{tab:disclosure_behavior_appendix}
\end{table}
\begin{table}
\centering
\begin{small}
\begin{tabular}{llccccc}
\toprule
Dimension & Family & Q1 & Q4 & Q1-Q4 & Q1 share & Q1 n \\
\midrule
Agency & Substitution & 0.47 [0.36, 0.57] & -0.43 [-0.53, -0.31] & 0.90 [0.74, 1.03] & 44.2\% & 405 \\
Agency & Pruning & 0.42 [0.25, 0.58] & -0.33 [-0.46, -0.20] & 0.75 [0.54, 0.95] & 25.0\% & 229 \\
Agency & Expansion & 0.33 [0.15, 0.50] & -0.25 [-0.38, -0.13] & 0.58 [0.36, 0.78] & 26.7\% & 245 \\
Agency & Restructuring & 0.32 [0.00, 0.60] & -0.11 [-0.35, 0.11] & 0.43 [0.03, 0.85] & 4.1\% & 38 \\
\midrule
Economy & Pruning & 0.43 [0.24, 0.60] & -0.27 [-0.39, -0.17] & 0.69 [0.48, 0.91] & 25.0\% & 229 \\
Economy & Expansion & 0.40 [0.19, 0.62] & -0.31 [-0.43, -0.22] & 0.72 [0.49, 0.97] & 25.5\% & 234 \\
Economy & Substitution & 0.31 [0.21, 0.41] & -0.37 [-0.47, -0.28] & 0.67 [0.53, 0.82] & 43.5\% & 399 \\
Economy & Restructuring & -0.12 [-0.37, 0.09] & -0.12 [-0.23, -0.04] & 0.01 [-0.28, 0.26] & 6.0\% & 55 \\
\midrule
Structure & Pruning & 0.55 [0.37, 0.70] & -0.35 [-0.47, -0.26] & 0.90 [0.71, 1.08] & 28.4\% & 260 \\
Structure & Substitution & 0.49 [0.36, 0.62] & -0.30 [-0.36, -0.24] & 0.79 [0.66, 0.93] & 41.5\% & 381 \\
Structure & Expansion & 0.24 [0.09, 0.40] & -0.43 [-0.55, -0.30] & 0.67 [0.48, 0.85] & 25.0\% & 229 \\
Structure & Restructuring & 0.18 [-0.38, 0.64] & -0.24 [-0.45, -0.06] & 0.42 [-0.17, 0.91] & 5.1\% & 47 \\
\midrule
Coherence & Substitution & 0.48 [0.34, 0.65] & -0.33 [-0.45, -0.21] & 0.81 [0.62, 1.02] & 42.0\% & 385 \\
Coherence & Expansion & 0.17 [0.04, 0.29] & -0.18 [-0.30, -0.07] & 0.36 [0.18, 0.51] & 28.4\% & 260 \\
Coherence & Pruning & 0.14 [0.06, 0.27] & -0.11 [-0.22, -0.03] & 0.26 [0.14, 0.43] & 25.5\% & 234 \\
Coherence & Restructuring & -0.01 [-0.27, 0.16] & -0.06 [-0.24, 0.07] & 0.05 [-0.24, 0.28] & 4.1\% & 38 \\
\midrule
Framing & Substitution & 0.31 [0.21, 0.44] & -0.38 [-0.57, -0.23] & 0.70 [0.53, 0.92] & 43.3\% & 400 \\
Framing & Expansion & 0.12 [0.04, 0.21] & -0.10 [-0.23, 0.03] & 0.21 [0.06, 0.36] & 25.9\% & 239 \\
Framing & Pruning & 0.07 [0.03, 0.13] & -0.18 [-0.35, -0.05] & 0.25 [0.12, 0.43] & 25.8\% & 238 \\
Framing & Restructuring & 0.05 [0.01, 0.18] & -0.23 [-0.63, 0.07] & 0.28 [-0.03, 0.72] & 5.0\% & 46 \\
\bottomrule

\end{tabular}
\end{small}
\caption{Behavior efficacy at the first and last quartile. Mean and 95\% confidence interval are reported for each quartile.}
\label{tab:behavior_ling_q1_q4}
\end{table}
\begin{table}[tb]
\centering
\begin{tabular}{l c c c c c}
\toprule
Dimension & AI $M (SD)$ & Human $M (SD)$ & $d_z$ & $t(44)$ & $p$ \\
\midrule
Agency & 0.10 (0.36) & -0.14 (0.46) & 0.45 & 2.99 & .008* \\
Economy & -0.06 (0.29) & -0.00 (0.32) & -0.20 & -1.32 & .242 \\
Structure & 0.13 (0.20) & -0.03 (0.30) & 0.54 & 3.62 & .004* \\
Coherence & -0.08 (0.41) & 0.17 (0.43) & -0.47 & -3.15 & .007* \\
Framing & 0.12 (0.54) & 0.15 (0.51) &  -0.04 & -0.26 & .796 \\
\bottomrule
\end{tabular}
\caption{Human vs AI initial linguistic properties.}
\label{tab:initial_states}
\end{table}
\begin{table}[htpb]
\centering
\begin{tabular}{l c c c c c}
\toprule
Dimension & AI $M (SD)$ & Human $M (SD)$ & $d_z$ & $t(44)$ & $p$ \\
\midrule
Agency & 0.01 (0.31) & -0.19 (0.38) & 0.46 & 3.07 & .018* \\
Economy & -0.02 (0.25) & 0.02 (0.27) & -0.17 & -1.17 & .309 \\
Structure & 0.11 (0.18) & 0.00 (0.24) & 0.41 & 2.76 & .019* \\
Coherence & -0.05 (0.31) & 0.13 (0.42) & -0.39 & -2.63 & .019* \\
Framing & 0.13 (0.50) & 0.11 (0.39) & 0.03 & 0.18 & .855 \\
\bottomrule
\end{tabular}
\caption{Human vs AI post-edit linguistic properties.}
\label{tab:post_states}
\end{table}
\begin{table}
\begin{tabular}{llcccccc}
\toprule
Dimension & Source & Pre $M$ (SD) & Post $M$ (SD) & Mean Change & $d_z$ & $t(44)$ & $p_{\mathrm{FDR}}$ \\
\midrule
Agency & AI & 0.10 (0.36) & -0.01 (0.30) & -0.11 & -0.60 & -4.04 & .001* \\
Agency & Human & -0.14 (0.46) & -0.16 (0.38) & -0.02 & -0.10 & -0.70 & .610 \\
\midrule
Economy & AI & -0.06 (0.29) & -0.02 (0.25) & 0.04 & 0.46 & 3.07 & .009* \\
Economy & Human & -0.00 (0.32) & 0.02 (0.26) & 0.02 & 0.24 & 1.61 & .303 \\
\midrule
Structure & AI & 0.13 (0.20) & 0.10 (0.17) & -0.03 & -0.32 & -2.16 & .060 \\
Structure & Human & -0.03 (0.30) & 0.00 (0.24) & 0.03 & 0.24 & 1.58 & .303 \\
\midrule
Coherence & AI & -0.08 (0.41) & -0.04 (0.30) & 0.04 & 0.22 & 1.49 & .178 \\
Coherence & Human & 0.17 (0.43) & 0.16 (0.37) & -0.01 & -0.06 & -0.40 & .695 \\
\midrule
Framing & AI & 0.12 (0.54) & 0.14 (0.48) & 0.02 & 0.14 & 0.95 & .346 \\
Framing & Human & 0.15 (0.51) & 0.11 (0.37) & -0.04 & -0.16 & -1.09 & .470 \\
\bottomrule
\end{tabular}
\caption{Original and edited abstract differences according to a paired t-test.}
\label{tab:app_delta_abs}
\end{table}

\section{Additional details on Language Models as editing assistants}
\label{app:lm_exp}

Additional results relating to \Cref{tab:app_lm_zeroshot,tab:app_lm_zeroshot_comp,tab:app_delta_zeroshot_percentile,tab:rubric_vs_scoped,tab:app_scope_comparison,tab:app_burst_prepost}.

We use standard decoding parameters: temperature $1.0$ and $\text{top}_\text{p}$ $0.95$ for all models. The specific versions are \texttt{gpt-5.4-2026-03-05}, \texttt{claude-opus-4-6}, and \texttt{gemini-3.1-pro}.

\subsection{Regarding LM choices and differences.}

\begin{figure}[t]
  \centering

  \begin{subfigure}{0.95\textwidth}
    \centering
    \includegraphics[width=\textwidth]{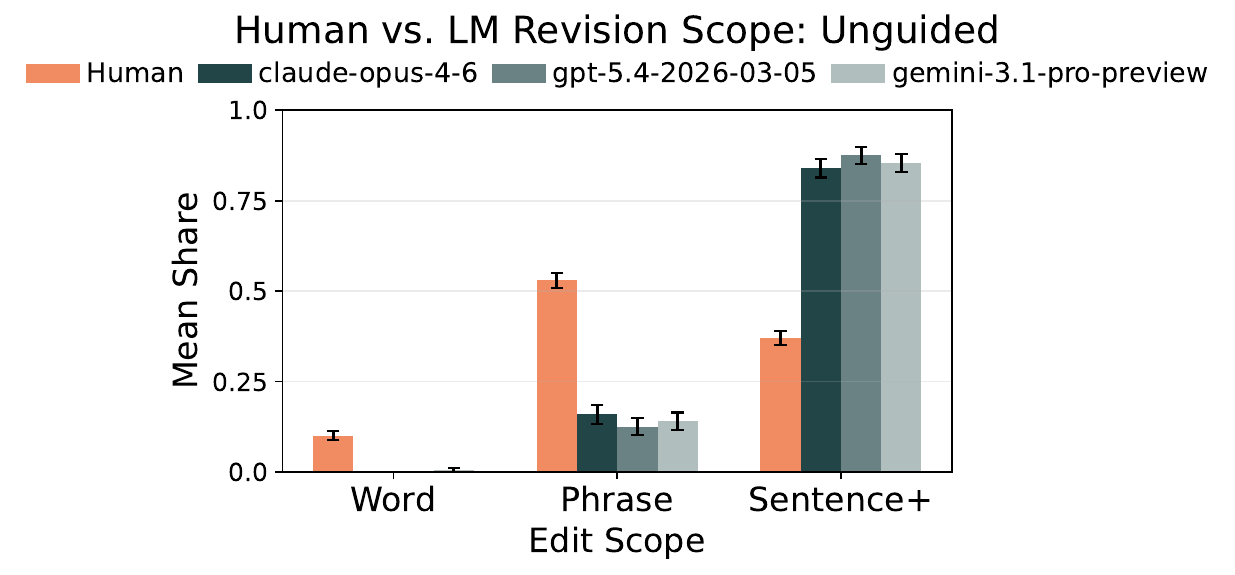}
    \caption{Unguided}
    \label{fig:app_align_unguided}
  \end{subfigure}

  \vspace{0.5em}

  \begin{subfigure}{0.95\textwidth}
    \centering
    \includegraphics[width=\textwidth]{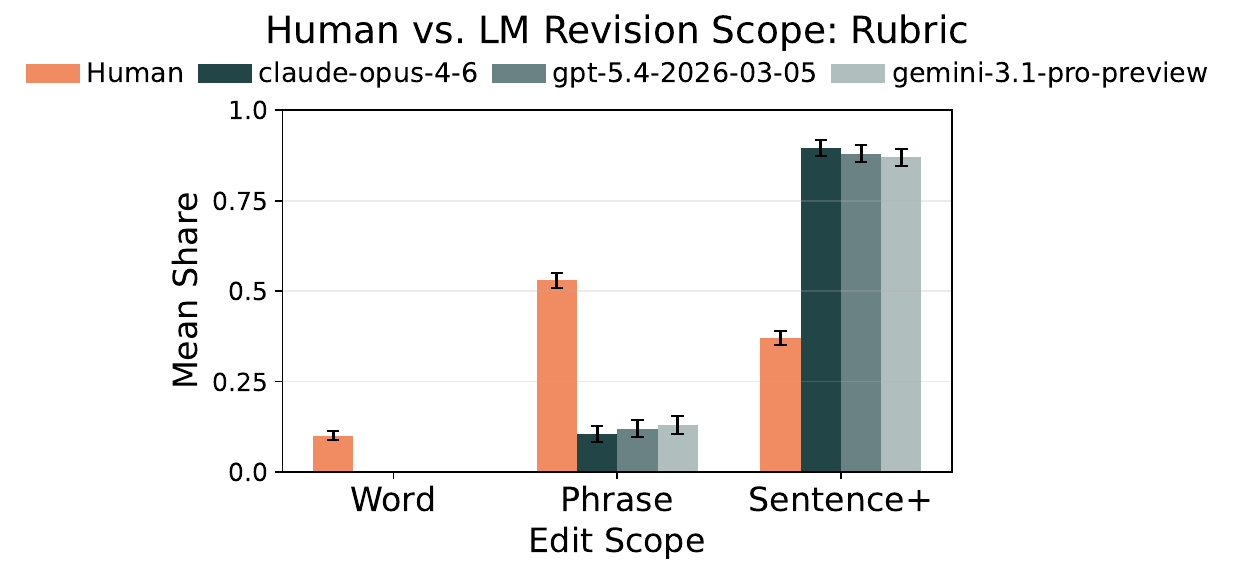}
    \caption{Rubric}
    \label{fig:app_align_rubric}
  \end{subfigure}

  \vspace{0.5em}

  \begin{subfigure}{0.95\textwidth}
    \centering
    \includegraphics[width=\textwidth]{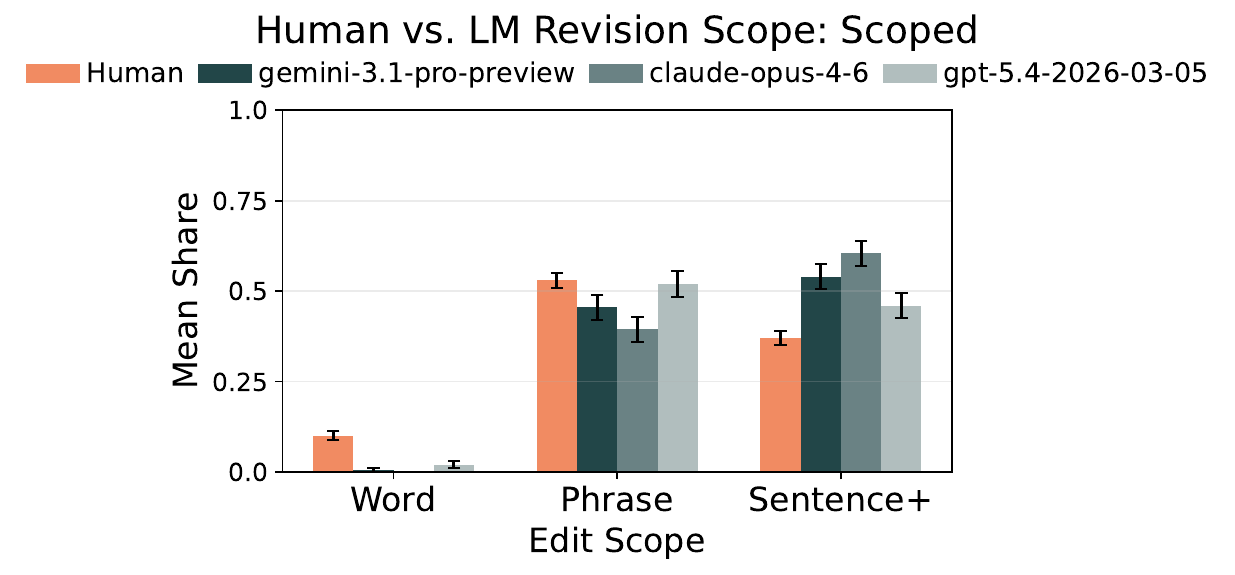}
    \caption{Scoped}
    \label{fig:app_align_scoped}
  \end{subfigure}

  \caption{Scope alignment by model across prompting conditions.}
  \label{fig:app_align}
\end{figure}

\begin{figure}[t]
    \centering
    \includegraphics[width=\linewidth]{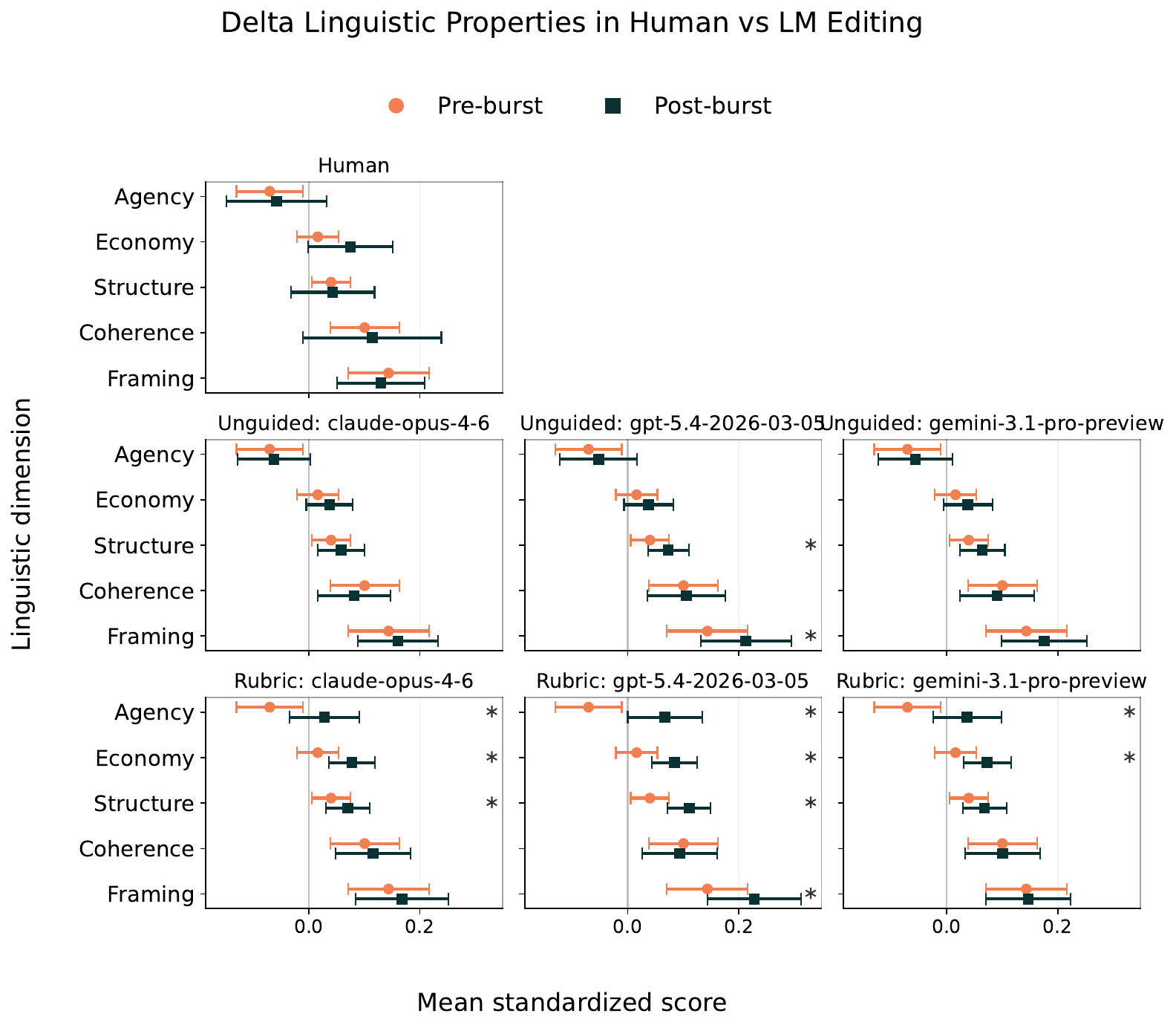}
    \caption{Linguistic properties of original and edited abstracts. In this case, the editor and prompt vary per plot.}
    \label{fig:app_prepost_model}
\end{figure}

\label{app:lm_exp_diff}
ChatGPT and Claude are the two systems real users are most likely to use. We additionally add Gemini 3.1 Pro as a comparator. We ran the main analyses for each model independently and report them below. They largely follow the same trend, consistent with our main takeaways.

In the editing-scope experiments, models show roughly similar changes to one another (see \autoref{fig:app_align}).

We separate edit-burst improvements by model across the experimental categories. Similar to the combined model scores, each makes more consistent improvements to sentence-level metrics, but none improve coherence (see \autoref{fig:app_prepost_model}).

The minor perturbations in performance do not form any convincing narratives. However, our reported results should be interpreted within the scope of the respective models and are not generalizable.

\begin{table}[tb]
\centering

\begin{tabular}{llccc}
\toprule
Dimension & Editor & Mean Diff & Cohen's $d_z$ & $p_\text{FDR}$ \\
\midrule
Agency & Human & -0.001 & -0.01 & 0.918 \\
Agency & LM: Unguided & 0.056 & 0.15 & 0.137 \\
Agency & LM: Rubric & 0.215 & 0.50 & $<.001$ \\
\midrule
Economy & Human & 0.005 & 0.05 & 0.320 \\
Economy & LM: Unguided & 0.053 & 0.31 & $<.001$ \\
Economy & LM: Rubric & 0.111 & 0.53 & $<.001$ \\
\midrule
Structure & Human & -0.003 & -0.03 & 0.633 \\
Structure & LM: Unguided & 0.017 & 0.08 & 0.428 \\
Structure & LM: Rubric & 0.086 & 0.37 & $<.001$ \\
\midrule
Coherence & Human & 0.018 & 0.06 & 0.223 \\
Coherence & LM: Unguided & -0.048 & -0.13 & 0.167 \\
Coherence & LM: Rubric & 0.002 & 0.00 & 0.952 \\
\midrule
Framing & Human & 0.006 & 0.03 & 0.618 \\
Framing & LM: Unguided & 0.127 & 0.22 & 0.016 \\
Framing & LM: Rubric & 0.104 & 0.17 & 0.081 \\
\bottomrule
\end{tabular}
\caption{LM as a zero-shot editor paired t-test.}
\label{tab:app_lm_zeroshot}
\end{table}
\begin{table}[tb]
\centering

\begin{tabular}{l c c c c c}
\toprule
Dimension & Editor 1 & Editor 2 & Mean Diff & {$\beta$} & {$p$} \\
\midrule
Agency & LM: Rubric & Human & 0.216 & 0.836 & $<.001$ \\
Economy & LM: Rubric & Human & 0.107 & 0.816 & $<.001$ \\
Structure & LM: Rubric & Human & 0.094 & 0.618 & $<.001$ \\
Coherence & LM: Rubric & Human & -0.027 & -0.088 & .811 \\
Framing & LM: Rubric & Human & 0.099 & 0.286 & .021 \\
\midrule
Agency & LM: Rubric & LM: Unguided & 0.158 & 0.392 & $<.001$ \\
Economy & LM: Rubric & LM: Unguided & 0.055 & 0.276 & .007 \\
Structure & LM: Rubric & LM: Unguided & 0.069 & 0.308 & .003 \\
Coherence & LM: Rubric & LM: Unguided & 0.046 & 0.122 & .154 \\
Framing & LM: Rubric & LM: Unguided & -0.018 & -0.031 & .686 \\

\bottomrule
\end{tabular}
\caption{LM as a zero-shot editor comparative results from a Welch's t-test.}
\label{tab:app_lm_zeroshot_comp}
\end{table}
\begin{table}
\begin{tabular}{lllrrrrrrrl}
\toprule
Dimension & Editor & Percentile & Pre mean & Post mean & $\Delta$ mean  & t & {$p_\text{FDR}$} \\
\midrule
Agency & LM: Rubric & Below median & 25.69 & 60.15 & 34.46 & 11.48 & $<.001$ \\
Agency & LM: Rubric & Above median & 74.94 & 71.28 & -3.65 & -1.65 & 0.114 \\
\midrule
Economy & LM: Rubric & Below median & 24.44 & 42.49 & 18.05 & 7.51 & $<.001$ \\
Economy & LM: Rubric & Above median & 74.04 & 78.81 & 4.77 & 2.11 & 0.048 \\
\midrule
Structure & LM: Rubric & Below median & 25.79 & 45.26 & 19.48 & 6.53 & $<.001$ \\
Structure & LM: Rubric & Above median & 75.94 & 73.29 & -2.66 & -1.03 & 0.307 \\
\midrule
Coherence & LM: Rubric & Below median & 25.18 & 35.65 & 10.47 & 3.82 & $<.001$ \\
Coherence & LM: Rubric & Above median & 73.78 & 59.85 & -13.92 & -5.28 & $<.001$ \\
\midrule
Framing & LM: Rubric & Below median & 33.01 & 50.14 & 17.13 & 6.70 & $<.001$ \\
Framing & LM: Rubric & Above median & 76.02 & 63.96 & -12.06 & -4.20 & $<.001$ \\
\bottomrule
\end{tabular}
\caption{Paired t-test statistics for pre and post edits, segmented into high and low percentiles. Corresponds to \autoref{fig:lm_prepost_percentiles}.}
\label{tab:app_delta_zeroshot_percentile}
\end{table}
\begin{table}
\begin{tabular}{llccclll}
\toprule
Dimension & Initial group & Rubric $\Delta$ & Scoped $\Delta$ & $\Delta$ improvement & 95\% CI & $p_{\mathrm{FDR}}$ \\
\midrule
Agency & Low initial & 12.89 & 6.34 & -6.55 & [-9.95, -3.14] & 0.002 \\
Agency & High initial & 2.39 & 2.00 & -0.39 & [-2.44, 1.66] & 0.788 \\
\midrule
Economy & Low initial & 9.36 & 4.28 & -5.08 & [-8.05, -2.10] & 0.004 \\
Economy & High initial & 2.81 & 2.87 & 0.06 & [-2.38, 2.50] & 0.962 \\
\midrule
Structure & Low initial & 11.40 & 5.67 & -5.73 & [-9.36, -2.09] & 0.007 \\
Structure & High initial & -2.55 & -3.46 & -0.92 & [-3.89, 2.06] & 0.682 \\
\midrule
Coherence & Low initial & 5.33 & 3.50 & -1.83 & [-4.70, 1.04] & 0.351 \\
Coherence & High initial & -5.30 & -2.79 & 2.51 & [-0.60, 5.62] & 0.228 \\
\midrule
Framing & Low initial & 6.68 & 2.86 & -3.82 & [-6.57, -1.08] & 0.016 \\
Framing & High initial & -2.49 & -1.61 & 0.88 & [-1.62, 3.37] & 0.682 \\
\bottomrule
\end{tabular}
\caption{Direct comparison of percentile improvement under rubric-guided versus scoped prompting. Negative values in $\Delta$ improvement indicate smaller gains under scoped prompting.}
\label{tab:rubric_vs_scoped}
\end{table}
\begin{table}
\centering
\begin{tabular}{llccccc}
\toprule
Experiment & Scope & Human & LM & $\Delta$LM--Human & $d_z$ & $p_{\text{FDR}}$ \\
\midrule
Unguided & Word & 0.10 & 0.00 & -0.10 & -0.33 & $<.001$ \\
Rubric & Word & 0.10 & 0.00 & -0.10 & -0.33 & $<.001$ \\
Scoped & Word & 0.10 & 0.01 & -0.09 & -0.28 & $<.001$ \\
Unguided & Phrase & 0.53 & 0.14 & -0.39 & -0.68 & $<.001$\\
Rubric & Phrase & 0.53 & 0.11 & -0.42 & -0.78 & $<.001$ \\
Scoped & Phrase & 0.53 & 0.46 & -0.07 & -0.14 & $<.001$ \\
Unguided & Sentence+ & 0.37 & 0.86 & 0.49 & 0.93 & $<.001$ \\
Rubric & Sentence+ & 0.37 & 0.89 & 0.52 & 1.03 & $<.001$ \\
Scoped & Sentence+ & 0.37 & 0.53 & 0.16 & 0.39 & $<.001$ \\
\bottomrule
\end{tabular}
\caption{Human and LM burst scope distributions. Positive values indicate the LM uses the corresponding scope more often than humans.}
\label{tab:app_scope_comparison}
\end{table}
\begin{table}
\centering
\begin{tabular}{llccc}
\toprule
Editor & Dimension & Mean change & $d_z$ & $p_\text{FDR}$ \\
\midrule
Human & Agency & 0.01 & 0.02 & 0.866 \\
Human & Economy & 0.06 & 0.11 & 0.207 \\
Human & Structure & 0.00 & 0.01 & 0.930 \\
Human & Coherence & 0.01 & 0.02 & 0.866 \\
Human & Framing & -0.01 & -0.05 & 0.722 \\
\midrule
Unguided & Agency & 0.01 & 0.07 & 0.247 \\
Unguided & Economy & 0.02 & 0.13 & 0.020 \\
Unguided & Structure & 0.03 & 0.15 & 0.007 \\
Unguided & Coherence & -0.01 & -0.03 & 0.772 \\
Unguided & Framing & 0.04 & 0.19 & $<.001$ \\
\midrule
Rubric & Agency & 0.12 & 0.48 & $<.001$ \\
Rubric & Economy & 0.06 & 0.38 & $<.001$ \\
Rubric & Structure & 0.05 & 0.27 & $<.001$ \\
Rubric & Coherence & 0.00 & 0.01 & 0.866 \\
Rubric & Framing & 0.05 & 0.14 & 0.0109 \\
\bottomrule
\end{tabular}
\caption{Paired t-tests of burst-level linguistic scores for human and LM editors.}
\label{tab:app_burst_prepost}
\end{table}

\newpage
\clearpage

\subsection{Prompts}
\label{app:prompts}

\FloatBarrier
\begin{takeawaybox2}
\small
\textbf{Unguided}

\# Goal

Your goal is to improve the quality of the scientific abstract.

\#\# What is an abstract?

An abstract is a short summary of completed research. It is intended to describe the work without going into great detail. Abstracts should be self-contained and concise, explaining your work as briefly and clearly as possible. An abstract should be able to stand independently from the research paper and still tell the reader something significant. The most important function of an abstract is to help a reader decide if he or she is interested in reading your entire publication.

\#\# An effective abstract will contain several key features:

Motivation or problem statement: Why is the research/argument important? What practical, scientific, theoretical or artistic gap is the project filling?
\\ \\
Methods/procedure/approach: What did the researcher actually do to get your results? (e.g. analyzed 3 novels, completed a series of 5 oil paintings, interviewed 17 students)
\\ \\
Results/findings/product: After completing the above procedure, what did the researcher learn/invent/create?
\\ \\
Conclusion/implications: What are the larger implications of the findings, especially for the problem/gap identified previously? Why is this research valuable?
\\ \\
Keep the abstract short: A general rule of abstract length is 150-200 words.
\\ \\
Do not add any new information: If something doesn't appear in the input, then don't put it in the abstract. An abstract is supposed to convey scientific findings, so they have to be precise and factual. Please don't embellish any results or findings.
\\ \\
\#\# Original Abstract
Below is the original abstract before editing.
\\ \\
\{abstract\}
\\ \\
\#\# Output Format

Please think step by step to complete your goal, then provide the abstract in full. Your response should be in the following JSON format:

\smallskip

\begin{verbatim}
{
  "reasoning": "the reasoning here",
  "edited_abstract": "the full edited abstract here"
}
\end{verbatim}
\end{takeawaybox2}

\FloatBarrier
\begin{takeawaybox2}
\textbf{Rubric}
\small

\textbf{Goal}\par

Your goal is to improve the quality of the scientific abstract.\par

\textbf{What is an abstract?}\par

An abstract is a short summary of completed research. It is intended to describe the work without going into great detail. Abstracts should be self-contained and concise, explaining your work as briefly and clearly as possible. An abstract should be able to stand independently from the research paper and still tell the reader something significant. The most important function of an abstract is to help a reader decide if he or she is interested in reading your entire publication.\par

\textbf{An effective abstract will contain several key features:}\par

\textit{Motivation or problem statement:} Why is the research/argument important? What practical, scientific, theoretical or artistic gap is the project filling?\par

\textit{Methods/procedure/approach:} What did the researcher actually do to get your results? (e.g., analyzed 3 novels, completed a series of 5 oil paintings, interviewed 17 students)\par

\textit{Results/findings/product:} After completing the above procedure, what did the researcher learn/invent/create?\par

\textit{Conclusion/implications:} What are the larger implications of the findings, especially for the problem/gap identified previously? Why is this research valuable?\par

Keep the abstract short: A general rule of abstract length is 150--200 words.\par

Do not add any new information: If something doesn't appear in the input, then don't put it in the abstract. An abstract is supposed to convey scientific findings, so they have to be precise and factual. Please don't embellish any results or findings.\par

\medskip
\textbf{Rubric for Writing}\par

\textbf{Goal}\par

Your goal is to improve the quality of the scientific abstract across five dimensions: Agency, Economy, Structure, Coherence, and Framing.\par

\textbf{Metric Rubric}\par

Each dimension is scored on a standardized scale relative to a reference distribution of typical pre-edit scientific writing:
\begin{itemize}
\item 0 = typical
\item positive = better than typical
\item negative = worse than typical
\end{itemize}

\textbf{Agency}\par
Definition: clarifies the role between the actors (subjects) and their actions (verbs). Clear writing makes it easy to identify who is doing what.

\begin{itemize}
\item lexical-verb ratio: proportion of semantically meaningful verbs relative to all finite verbs.
\item passive-voice share: the number of sentences that use passive constructions.
\item character-as-subject: how often the grammatical subject refers to a real actor.
\item agentless-passive: how often passive clauses omit a by-phrase agent.
\item metadiscourse we-density: how often writers use first-person pronouns to introduce claims.
\end{itemize}

\textbf{Economy}\par
Definition: quantifies how efficiently a text conveys complex information without unnecessary compression.

\begin{itemize}
\item nominalization density: Measures how often verbs and adjectives are turned into abstract nouns.
\item abstract-congestion: Detects “congested” sentences where an abstract-noun subject is accompanied by additional nominalizations.
\item compound-noun rate: Captures the number of long noun strings (three or more consecutive nouns).
\item prep.\ phrase density: Counts of in for relations per 100 words.
\item filler-word density: Counts empty hedges and intensifiers like very, actually, basically.
\end{itemize}

\textbf{Structure}\par
Definition: describes how the arrangement of words and phrases leads to the primary grammatical relationships.

\begin{itemize}
\item subject-onset distance: Measures how many tokens occur before the main grammatical subject.
\item initial-momentum: Metric gauges how soon the main verb appears relative to the sentence start.
\item subject-verb gap: Measures the number of words between a subject and its verb.
\item verb-object gap: Captures how many tokens separate a verb from its direct object.
\item post-verb density: Counts subordinate clauses or heavy additions that follow the main verb.
\end{itemize}

\textbf{Coherence}\par
Definition: measures how well a text introduces new information and connects threads.

\begin{itemize}
\item cohesion-chain strength: Measures how well consecutive sentences connect via repeated or echoed content words.
\item old-to-new ordering: Calculates the relative position of previously unseen words in each sentence.
\item topic-continuity: Checks whether successive sentences share the same grammatical subject.
\item stress-position rate: Measures how often the stress position is occupied by a content-bearing word.
\end{itemize}

\textbf{Framing}\par
Definition: tracks the placement of context and claims.

\begin{itemize}
\item pivot-explicitness: Checks for clear adversative markers (but, however, yet) signaling a shift.
\item condition-and-cost: Examines whether problem statements articulate both the condition and the consequence.
\item claim placement: Captures where the main contribution or result (e.g., "we show") appears.
\end{itemize}

\medskip
\textbf{Original Abstract}\par

\medskip
\textbf{Output Format}\par

Please think step by step to complete your goal, then provide the abstract in full. Your response should be in the following JSON format:

\begin{verbatim}
   {
  "reasoning": "<the reasoning here>",
  "edited_abstract": "<the full edited abstract here>"
} 
\end{verbatim}
\end{takeawaybox2}

\end{document}